\documentclass[journal]{IEEEtran}

\usepackage{amssymb,mathrsfs,color}
\usepackage[cmex10]{amsmath}
\usepackage{array}
\usepackage{graphicx, color}
\usepackage{latexsym}
\usepackage{boxedminipage}
\usepackage{fancybox}
\usepackage{ascmac}
\usepackage{algorithm}
\usepackage{algorithmic}
\usepackage{amsmath}
\usepackage{cite}

\newtheorem{theorem}         {Theorem}

\newtheorem{lemma}       [theorem]      {Lemma}

\newtheorem{proposition}       [theorem]      {Proposition}

\newtheorem{problem}{Problem\hspace{-1.0mm}}

\begin{document}
%
\title{
Transient Response Improvement for Interconnected Linear Systems:\\
Low-Dimensional Controller Retrofit Approach
}

%
%
%

\author{
Takayuki~Ishizaki,~\IEEEmembership{Member,~IEEE,}
Masakazu~Koike,~\IEEEmembership{Member,~IEEE,}
Jun-ichi~Imura,~\IEEEmembership{Member,~IEEE.}
\thanks{T.~Ishizaki and J.~Imura are with Graduate School of Engineering, Tokyo Institute of Technology, 2-12-1, Ookayama, Meguro, Tokyo, 152-8552, Japan.
e-mail: \{ishizaki,imura\}@mei.titech.ac.jp.}
\thanks{M.~Koike is with Tokyo University of Marine Science and Technology; 4-5-7, Kounan, Minato, Tokyo, 108-0075, Japan.
e-mail: mkoike0@kaiyodai.ac.jp.}
\thanks{This research was supported by CREST, JST.}
}

%
%

\markboth{Preprint}%
{Shell \MakeLowercase{\textit{et al.}}: Bare Demo of IEEEtran.cls for Journals}
%



\maketitle

\begin{abstract}
In this paper, we propose a method of designing low-dimensional retrofit controllers for interconnected linear systems.
In the proposed method, by retrofitting an additional low-dimensional controller to a preexisting control system, we aim at improving transient responses caused by spatially local state deflections, which can be regarded as a local fault occurring at a specific subsystem.
It is found that a type of state-space expansion, called hierarchical state-space expansion, is the key to systematically designing a low-dimensional retrofit controller, whose action is specialized to controlling the corresponding subsystem.
Furthermore, the state-space expansion enables theoretical clarification of the fact that the performance index of the transient response control is improved by appropriately tuning the retrofit controller.
The efficiency of the proposed method is shown through a motivating example of power system control where we clarify the trade-off relation between the dimension of  a retrofit controller and its control performance.
\end{abstract}



%
\IEEEpeerreviewmaketitle

%
%
%
%

\section{Introduction}

\IEEEPARstart{M}{ost} existing control systems consist of different kinds of physical and artificial dynamical components.
For example, in the frequency control of power systems \cite{kundur1994power}, several types of centralized control strategies, called load frequency control and economic dispatch control, are implemented to stabilize frequency variations  over a few dozen minutes, while those over a few minutes are stabilized by the inherent stability of physical appliances, i.e., the governor droop control.
To develop such a practically working control system, it would be desirable to make control systems retrofittable in the sense that additional controllers can be retrofitted to accomplish individual objectives in a distributed fashion.
In particular, it is desirable that the additional controllers can be designed independently of  preexisting controllers.
Such a property is relevant to preventing the redesign of additional controllers when the modification of preexisting controllers is required.

With this background, this paper aims at developing a method of designing low-dimensional retrofit controllers for interconnected linear systems.
By retrofitting a low-dimensional local controller to a preexisting control system using the proposed method, we seek to improve transient responses caused by spatially local state deflections, which can be regarded as possible contingencies in a specific area.
This type of retrofit control enables more tractable handling of local contingencies, such as unexpected disturbances, in the sense that each of the spatially local state deflections can be suppressed in a distributed fashion.
In fact, the notion of intelligent Balancing Authorities \cite{baros2014intelligent}, which corresponds to the portfolios of appliances responsible for local area control, is introduced to suppress the backbone propagation of faults in power systems control.
The proposed retrofit controller is designed such that the entire closed-loop system with the retrofit controller is stable for any preexisting controller that stabilizes the preexisting control system.
Furthermore, we guarantee the improved transient responses of the entire closed-loop system as improving the performance of a low-dimensional local controller, which is involved in the retrofit controller.
These particular properties stem from the fact that the retrofit controller can be designed independently of the preexisting controller.

To design such a retrofit controller for transient response improvement in this paper, we utilize a type of state-space expansion, called \textit{hierarchical state-space expansion} \cite{sadamoto2014hierarchical,ishizaki2014hierarchicalEPJ}.
The hierarchical state-space expansion utilizes a projection-based model reduction technique \cite{antoulas2005approximation} to generate a redundant realization of interconnected systems having a cascade structure that enables the distributed design \cite{langbort2010distributed} of retrofit controllers.
On the basis of this state-space expansion, we theoretically show that the performance index of the transient response control is improved by appropriately tuning the retrofit controller whose dimension is considerably lower than that of the whole system to be controlled.

As a demonstration of the effectiveness of our retrofit control, we perform  numerical simulation of the transient stabilization in power systems control \cite{kundur1994power}.
In this simulation, while supposing the preexistence of a broadcast-type feedback controller with a sampling and holding time, which corresponds to a conventional controller for automatic generation control (AGC), we show that a low-dimensional local controller retrofitted to the preexisting control system can improve the transient response performance for a local fault at a specific generator.
Furthermore, we show the trade-off relation between the dimension of retrofit controllers and their control performance, with consideration of the allocation of input and output ports for retrofit control.

To clarify our contribution, some references regarding control system design based on additional compensation are in order.
In \cite{izumi2002improving}, a method to improve the transient response of control systems is proposed on the basis of compensation by an additional input signal and the selection of initial controller states.
In this method, by regarding the control system as an autonomous system, the compensation is applied in a feedforward manner;
thus, feedback control for unknown disturbances is not considered.
On the other hand, \cite{girard2009hierarchical} considers a hierarchical control architecture, where a low-dimensional model is used to construct an additional input signal such that the error between the output of the model and its original system converges to zero asymptotically.
However, a hierarchical control system is not necessarily easy to implement in practice because it assumes an exact model reduction, i.e., the low-dimensional model can exactly reconstruct the original system behavior the state feedback of the original system.

We give some references regarding control system design on multiple spatiotemporal scales.
From the viewpoint of time scale separation, we see a similarity between the proposed transient response control method and a control synthesis method based on singular perturbation theory \cite{kokotovic1972singular,ozguner1979near}.
In the singular perturbation-based approach, an asymptotic expansion is generally used to analyze the degradation of control performance due to the approximation.
By contrast, our approach has the advantage in that, on the basis of the hierarchical state-space  realization having a tractable cascade structure, we can analytically handle an approximation error of the low-dimensional model.
This redundant realization is different from those used in \cite{vsiljak2005control,ikeda1986overlapping} in the sense that we use state-space expansion to derive a cascade realization from the viewpoints of controllability and observability, whereas the existing works use it to approximately decouple interconnected systems only from the viewpoint of quasi block-diagonalization.
A preliminary version of this paper is found as \cite{ishizaki2015multiresolved}.
In comparison to that paper, this paper provides detailed proofs of the theoretical results, as well as analysis of the performance of the transient response control.

This paper is organized as follows.
In Section II, with a motivating example from AGC, we first formulate the design problem of retrofit controllers for  transient response improvement in a discrete-time setting.
Then, in Section III, after providing an overview of a control system design approach based on the hierarchical state-space expansion, we give a solution to the retrofit controller design problem.
In particular, we show that the performance index of transient response control is improved by suitably tuning the retrofit controller.
Section IV provides an example of the stabilization of frequency variations in a power network.
Finally, concluding remarks are provided in Section V.

\vspace{1.5pt}

\textit{Notation:}
We denote the set of real values by $\mathbb{R}$, the identity matrix by $I$, the all-ones vector by $\mbox{\boldmath ${\mathit 1}$}$, the image of a matrix $M$ by ${\rm im}$\hspace{1.5pt}$M$, the kernel by ${\rm ker}$\hspace{1.5pt}$M$, a left inverse of a left invertible matrix $P$ by $P^{\dagger}$, the finite-horizon and infinite-horizon $l_{2}$-norms of a square-summable vector sequence $f_{t}$ by 
\[
\|f_{t}\|_{l_{2}[T]}:=\sqrt{\textstyle{\sum_{t=0}^{T}\|f_{t}\|^{2}}},\quad
\|f_{t}\|_{l_{2}}:=\sqrt{\textstyle{\sum_{t=0}^{\infty}\|f_{t}\|^{2}}},
\]
the $h_{2}$-norm of a stable proper transfer matrix $G$ by 
\[\textstyle
\|G(z)\|_{h_{2}}:=\sqrt{\frac{1}{2\pi}\int_{0}^{2\pi}
\|G(e^{j\theta})\|_{\rm F}^{2}d\theta},
\]
where $\|\cdot\|_{\rm F}$ denotes the Frobenius norm, and the $h_{\infty}$-norm of a stable transfer matrix $G$ by
\[
\|G(z)\|_{h_{\infty}}:=\sup_{\theta\in[0,2\pi]}\|G(e^{j\theta})\|
\]
where $\|\cdot\|$ denotes the induced 2-norm.

With $\mathbb{N}=\{1,\ldots,N\}$, we denote the block-diagonal matrix having matrices 
$M_{i}$ for $i\in \mathbb{N}$ on its diagonal blocks by
\[
{\rm  diag}(M_{1},\ldots,M_{N})={\rm diag}(M_{i})_{i\in \mathbb{N}}.
\]
For an index set $\mathcal{I}$, we denote the matrix composed of the column vectors of $I$ associated with $\mathcal{I}$ by $e_{\mathcal{I}}$.
A map $\mathcal{F}$ is said to be a dynamical map if the triplet $(x_{t},u_{t},y_{t})$ with $y_{t}=\mathcal{F}(u_{t})$ solves a system of difference equations
\[
x_{t+1}=f(x_{t},u_{t}),\quad y_{t}=g(x_{t},u_{t})
\]
with some functions $f$ and $g$, and an initial value $x_{0}$.

\section{Problem Formulation}\label{secpf}

\begin{figure}[t]
\begin{center}
\includegraphics[width=85mm]{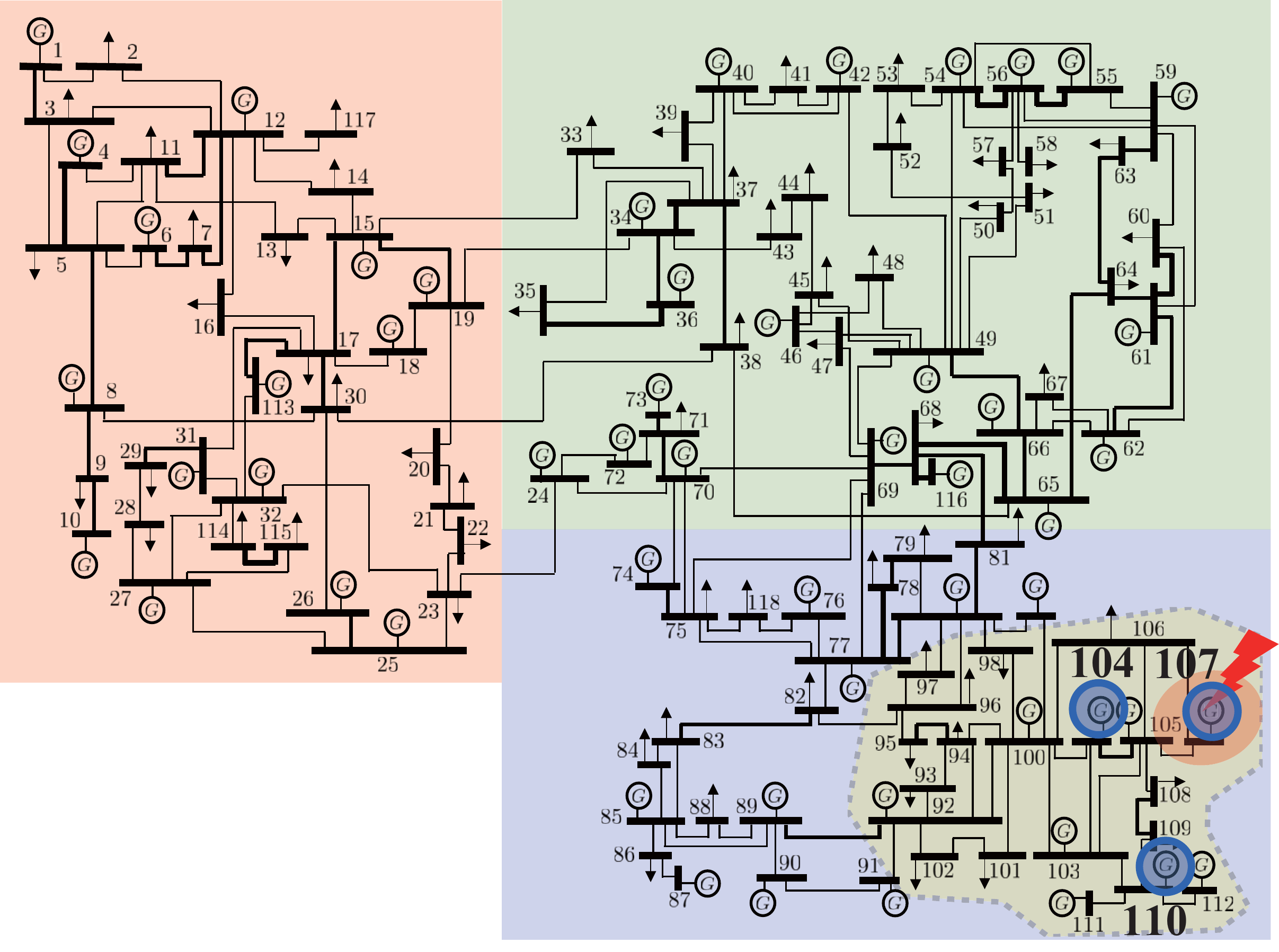}
\end{center}
\vspace{-3pt}
\caption{IEEE 118-bus test system composed of generators and loads.
Each generator is denoted by the symbol ``G'' and each load is denoted by the symbols ``$\uparrow$" or ``$\downarrow$".
An index is assigned to each of the buses.}
\label{figieee118}
\end{figure}

\begin{figure*}[t]
\begin{center}
\includegraphics[width=120mm]{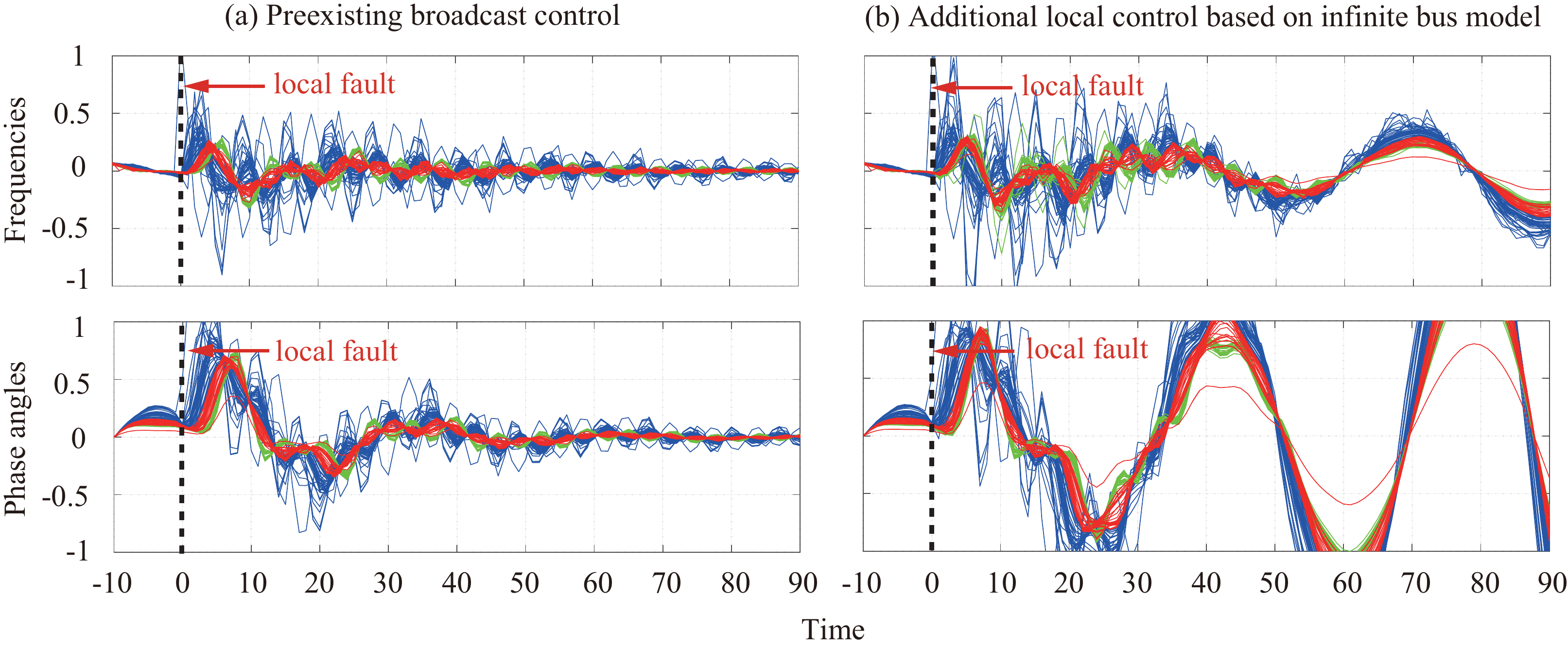}
\end{center}
\vspace{-6pt}
\caption{
System responses for local fault at generator on Bus 107.
The subfigures plot the frequencies and phase angles of all appliances whose line colors (blue, green, and red) are compatible with those of the appliance groups in Fig.~\ref{figieee118}.
}
\label{figbroad}
\end{figure*}

\subsection{Motivating Example from Automatic Generation Control}\label{secmot}

Let us consider a power system composed of 54 generators and 64 loads whose interconnection structure is given as the IEEE 118-bus test system shown in Fig.~\ref{figieee118}.
\begin{subequations}\label{eprsys}
With the label set of the generators denoted by $\mathcal{G}$, we suppose that the dynamics of each generator is described as a rotary appliance \cite{ilic1996hierarchical} of
\begin{equation}\label{rotar}
\dot{\theta}_{i}=\omega_{i},\quad m_{i}\dot{\omega}_{i}+d_{i}\omega_{i}+b_{i}u_{i}+e_{i}=0,\quad i\in \mathcal{G}
\end{equation}
where $\theta_{i}$ and $\omega_{i}$ denote the phase angle and the frequency, respectively, $e_{i}$ denotes the electric torque from other appliances, and $u_{i}$ denotes the mechanical torque input signal.
For the generators, the values of $m_{i}$ and $d_{i}$ are selected from the ranges of $[0.01,1]$ and $[0.007,0.01]$, respectively.
Furthermore, as the measurement output signals, we suppose that the phase angle $\theta_{i}$ and the frequency $\omega_{i}$ of all generators are measurable.

Similar to the generator dynamics, denoting the label set of the loads by $\mathcal{L}$, we describe each load dynamics as a rotary appliance of
\begin{equation}\label{rotarl}
\dot{\theta}_{i}=\omega_{i},\quad m_{i}\dot{\omega}_{i}+d_{i}\omega_{i}+e_{i}=0,\quad i\in \mathcal{L}
\end{equation}
for which we select the values of $m_{i}$ and $d_{i}$ from the same ranges as those for generators.
The interconnection among generators and loads can be represented as 
\begin{equation}
\textstyle
e_{i}=\sum_{j\in \mathcal{N}_{i}}Y_{i,j}(\theta_{j}-\theta_{i})
\end{equation}
where $\mathcal{N}_{i}$ denotes the index set associated with the neighborhood of the $i$th appliance and $Y_{i,j}$ denotes the admittance between the $i$th and $j$th appliances, the value of which is given in accordance with \cite{zimmerman2011matpower}.
\end{subequations}
The entire power system (\ref{eprsys}) is 236-dimensional.
In the following, we suppose that each state variable of generators and loads is defined as a deviation from desirable equilibria.

We consider a situation where a frequency control mechanism, called automatic generation control (AGC), has been implemented for the stabilization of the power system; see Section 9 in \cite{sauer1998power} for an overview of AGC.
For the stabilization, centralized secondary control is often implemented to an area composed of multiple generators and loads.
Let us suppose a zero-order hold input and a sampled output for controller implementation.
Then, the centralized secondary control can be represented as 
\begin{equation}\label{agcon}
\textstyle
u_{i}(\tau)=\kappa a_{i}\sum_{j\in \mathcal{G}}\omega_{j}(t\Delta t)+u_{i}^{\prime}(\tau),\quad\tau\in[t\Delta t,(t+1)\Delta t)
\end{equation}
where $\Delta t$ denotes the sampling period, $t$ denotes the time label, $\kappa$ denotes a feedback gain, $a_{i}$ denotes a scaling factor, and $u_{i}^{\prime}$ denotes another input signal injected to the feedback system.
In AGC, the aggregated frequency deviation $\sum_{j\in \mathcal{G}}\omega_{j}$ is called an area control error often denoted by ACE, and the scaling factor $a_{i}$ is called a participation factor, which is determined based on the level of contribution of the individual generators to the total generation control \cite{sauer1998power,green1996transformed}.
The feedback gain $\kappa$ is usually chosen in an empirical manner such that the resultant closed-loop system is stable.
Note that (\ref{agcon}) without $u_{i}^{\prime}$ can be seen as a broadcast-type controller.

It should be noted that the broadcast controller (\ref{agcon}) without $u_{i}^{\prime}$ cannot generally perform accurate control for individual generators because its input and output signals do not distinguish them.
In this sense, single use of the broadcast controller is not generally satisfactory for reducing the impact of a fault, such as a three-phase fault, at a particular bus.
We model such a local fault as an impulsive change in the initial phase angle of a particular generator.
More specifically, supposing that the fault occurs at the $\alpha$th generator that we focus on, we model it as
\begin{equation}\label{ifault}
(\theta_{\alpha}(0),\omega_{\alpha}(0))^{{\sf T}}=\delta_{0},\quad\alpha\in \mathcal{G}
\end{equation}
where $\delta_{0}\in \mathbb{R}^{2}$ is unknown.
In Fig.~\ref{figbroad}(a), we plot the transient state response supposing that a local fault occurs at the generator on Bus 107; see Fig.~\ref{figieee118} for its location.
The sampling period $\Delta t$ in (\ref{agcon}) is set as 1 [sec], the feedback gain $\kappa$ is set as 0.01, and $u_{i}^{\prime}$ is set as zero.
In this subfigure, the frequencies and phase angles of all appliances are plotted.
The colors of lines are associated with the colors of generator and load groups in Fig.~\ref{figieee118}.
From this figure, we see that the oscillation of frequencies and phase angles due to the local fault propagates to other appliance groups, shown by the red and green lines.

To suppress the propagation of the impact of the local fault, let us consider implementing an additional local controller that produces the input signal $u_{i}^{\prime}$ in (\ref{agcon}) using a couple of generators as a set of input and output ports.
Let us denote the label set of such input and output port generators by $\mathcal{J}_{\alpha}$.
For example, $\mathcal{J}_{\alpha}$ can be selected as the generators on Busses 104, 107, and 110 in Fig.~\ref{figieee118}, which are close to the particular generator that we focus on.
In the subsequent analysis, we assume that a label set $\mathcal{J}_{\alpha}$ is prespecified for the local fault (\ref{ifault}).

One simple approach to designing such an additional local controller is to apply a standard controller design technique, such as the linear quadratic regulator (LQR) design technique and the $\mathcal{H}_{2}$/$\mathcal{H}_{\infty}$-controller synthesis, where the feedback system composed of the power system (\ref{eprsys}) and the broadcast controller (\ref{agcon}) is regarded as a controlled plant.
\begin{subequations}\label{stlqr}
More specifically, representing $u_{i}^{\prime}$ in (\ref{agcon}) as
\begin{equation}\label{uprime}
u^{\prime}(\tau)=e_{\mathcal{J}_{\alpha}}\hat{u}(\tau)
\end{equation}
where $u^{\prime}\in \mathbb{R}^{|\mathcal{G}|}$ denotes the stacked vector of $u_{i}^{\prime}$, one can design a dynamical controller that produces $\hat{u}\in \mathbb{R}^{|\mathcal{J}_{\alpha}|}$ as
\begin{equation}\label{uhat}
\hat{u}(\tau)=\mathcal{K}_{\alpha}\left(e_{\mathcal{J}_{\alpha}}^{\sf T}\omega(t\Delta t)\right),\quad\tau\in[t\Delta t,(t+1)\Delta t)
\end{equation}
\end{subequations}
where $\omega$ denotes the stacked vector of $\omega_{i}$, and $\mathcal{K}_{\alpha}$ denotes the controller dynamical map.
The form of (\ref{uprime}) implies that $u_{i}^{\prime}=0$ holds for all generators such that $i\not\in \mathcal{J}_{\alpha}$, and (\ref{uhat}) implies that the dynamical controller feedbacks the sampled measurement of $(\omega_{i})_{j\in \mathcal{J}_{\alpha}}$.
This simple approach is, however, not necessarily reasonable for a large-scale power system because the dimension of the resultant controller is generally comparable with that of the preexisting control system of interest, i.e., the feedback system composed of the power system (\ref{eprsys}) and the broadcast controller (\ref{agcon}).
Moreover, modification of the broadcast controller, e.g., the tuning of the feedback gain $\kappa$, may impose the redesign of (\ref{stlqr}), because the feedback system involves (\ref{agcon}) as a system parameter.

In reality, a power system controller is often designed based on the model of an isolated area in a moderate size, where the remaining area is modeled as an infinite bus \cite{sauer1998power}.
Such an isolated area model can be regarded as a low-dimensional model of (\ref{eprsys}) obtained by neglecting the system properties of the remaining area and the broadcast controller (\ref{agcon}).
More specifically, let $\hat{\mathcal{G}}_{\alpha}\subset \mathcal{G}$ and $\hat{\mathcal{L}}_{\alpha}\subset \mathcal{L}$ denote the label sets of generators and loads belonging to the isolated area such that $\mathcal{J}_{\alpha}\subset\hat{\mathcal{G}}_{\alpha}$.
Then, we can describe the low-dimensional model as
\begin{subequations}\label{exred}
\begin{equation}
\begin{array}{ll}
\dot{\theta}_{i}=\omega_{i},\quad m_{i}\dot{\omega}_{i}+d_{i}\omega_{i}+b_{i}u_{i}^{\prime}+e_{i}=0,&\quad\hspace{-6pt}i\in\hat{\mathcal{G}}_{\alpha}\vspace{0pt}\\
\dot{\theta}_{i}=\omega_{i},\quad m_{i}\dot{\omega}_{i}+d_{i}\omega_{i}+e_{i}=0,&\quad\hspace{-6pt}i\in\hat{\mathcal{L}}_{\alpha}
\end{array}
\end{equation}
whose interconnection is represented as
\begin{equation}\textstyle
e_{i}=\sum_{j\in\hat{\mathcal{N}_{i}}}Y_{i,j}(\theta_{j}-\theta_{i}),\quad\hat{\mathcal{N}}_{i}:=\left(\mathcal{N}_{i}\cap(\hat{\mathcal{G}}_{\alpha}\cup\hat{\mathcal{L}}_{\alpha})\right).
\end{equation}
\end{subequations}
This low-dimensional model can represent, for example, the dynamics of the isolated area depicted by the dashed-line in Fig.~\ref{figieee118}, where the input and output port generators on Busses 104, 107, and 110 are involved.
The application of a standard controller design technique to such a low-dimensional model produces a low-dimensional controller in the form of (\ref{stlqr}).
However, the stability of the resultant closed-loop system is not ensured in general.
In fact, as shown in Fig.~\ref{figbroad}(b), a 36-dimensional local controller designed by the linear quadratic regulator (LQR) design technique induces the instability of the closed-loop system.
As demonstrated in this motivating example, systematic transient response improvement with a low-dimensional local controller is not generally straightforward.

\subsection{Description in General Discrete-Time Systems Form}

For the subsequent discussion, we describe the power system (\ref{eprsys}) with the broadcast controller (\ref{agcon}), which we call a preexisting controlled system, in a general discrete-time linear systems form.
Without loss of generality, we suppose that the generator label set $\mathcal{G}$ and the load label set $\mathcal{L}$ are given as
\[
\mathcal{G}=\{1,\ldots,N\},\quad\mathcal{L}=\{N+1,\ldots,N+M\}
\]
where $N:=|\mathcal{G}|$ and $M:=|\mathcal{L}|$ denote the numbers of generators and loads, respectively.
Furthermore, for convenience of discussion, we suppose that each load is associated with a generator, namely the load label set $\mathcal{L}$ is partitioned as
\[
\mathcal{L}=\mathcal{L}_{1}\cup\cdots\cup \mathcal{L}_{N},\quad\mathcal{L}_{i}\cap \mathcal{L}_{j}=\emptyset
\]
where each $\mathcal{L}_{i}$ is associated with a generator label $i\in \mathcal{G}$.
Note that any partition can be allowed without loss of generality in the following discussion.
Then, defining the state vector and the output signal of the $i$th subsystem as
\begin{equation}
x_{i}^{\rm c}:=\left(
(\theta_{i},\omega_{i})^{{\sf T}},(\theta_{j},\omega_{j})_{j\in \mathcal{L}_{i}}^{{\sf T}}
\right)^{{\sf T}},\quad
y_{i}^{\rm c}=\omega_{i},
\end{equation}
respectively,  we can represent the power system (\ref{eprsys}) as an interconnected system whose $i$th subsystem has the form of
\begin{equation}
\left\{
\begin{array}{ccl}
\dot{x}_{i}^{\rm c}&\hspace{-6pt}=&\hspace{-6pt}A_{ii}^{{\rm c}}x_{i}^{\rm c}+
\sum_{j\neq i}A_{ij}^{{\rm c}}x_{j}^{\rm c}+
B_{i}^{{\rm c}}u_{i}^{\rm c}\\
y_{i}^{\rm c}&\hspace{-6pt}=&\hspace{-6pt}C_{i}^{\rm c}x_{i}^{\rm c},
\end{array}
\right.
\end{equation}
which is $n_{i}:=2(1+|\mathcal{L}_{i}|)$-dimensional.
The system matrices $A_{ii}^{{\rm c}}$,  $A_{ij}^{{\rm c}}$,  $B_{i}^{{\rm c}}$, and $C_{i}^{\rm c}$ are determined in accordance with the system parameters and structures of (\ref{eprsys}), and the input signal $u_{i}$ in (\ref{rotar}) is rewritten by $u_{i}^{\rm c}$ for convenience.
The entire interconnected system is given as
\[
\left\{
\begin{array}{ccl}
\dot{x}^{\rm c}&\hspace{-6pt}=&\hspace{-6pt}A^{{\rm c}}x^{\rm c}+B^{{\rm c}}u^{\rm c}\\
y^{\rm c}&\hspace{-6pt}=&\hspace{-6pt}C^{\rm c}x^{\rm c}
\end{array}
\right.
\]
where $x^{\rm c}$,  $u^{\rm c}$, and $y^{\rm c}$ denote the stacked vectors of $x_{i}^{\rm c}$,  $u_{i}^{\rm c}$, and $y_{i}^{\rm c}$, respectively, $A^{\rm c}=(A_{ij}^{\rm c})_{i,j\in \mathcal{G}}$ and
\[
B^{{\rm c}}={\rm diag}(B_{i}^{{\rm c}})_{i\in \mathcal{G}},\quad C^{\rm c}={\rm diag}(C_{i}^{\rm c})_{i\in \mathcal{G}}.
\]

The local fault (\ref{ifault}) is represented as $x_{\alpha}^{\rm c}(0)=e_{1:2}\delta_{0}$, with $e_{1:2}\in \mathbb{R}^{n_{\alpha}\times 2}$ denoting the first and second column vectors of the $n_{\alpha}$-dimensional unit matrix.
This leads to the domain of initial conditions associated with the $\alpha$th subsystem of interest, denoted as
\begin{equation}\label{defclx}
\mathcal{X}_{\alpha}:=\left\{x^{\rm c}(0):x_{\alpha}^{\rm c}(0)=e_{1:2}\delta_{0},\ \delta_{0}\in \mathbb{R}^{2}\right\},
\end{equation}
which is assumed to be available.
Based on the discretization with the sampling period $\Delta t$ in (\ref{agcon}), we obtain the discrete-time system
\begin{equation}\label{sysd}
\Sigma:\left\{
\begin{array}{ccl}
x_{t+1}&\hspace{-6pt}=&\hspace{-6pt}Ax_{t}+Bu_{t}\\
y_{t}&\hspace{-6pt}=&\hspace{-6pt}Cx_{t},
\end{array}
\right.\quad x_{0}\in \mathcal{X}_{\alpha}
\end{equation}
where the system matrices are given as
\[\textstyle
A:=e^{A^{\rm c}\Delta t},\quad B:=\left(\int_{0}^{\Delta t}e^{A^{\rm c}t}dt\right)B^{{\rm c}},
\quad C:=C^{\rm c}.
\]
This satisfies $x^{\rm c}(t\Delta t)=x_{t}$ under the zero-order hold input
\[
u^{\rm c}(\tau)=u_{t},\quad\tau\in[t\Delta t,(t+1)\Delta t).
\]
We denote the dimension of $\Sigma$ by $n$.

For consistency with the composite input signal in (\ref{agcon}) with (\ref{uprime}), we describe the input signal $u_{t}$ in (\ref{sysd}) as
\begin{equation}\label{resin}
u_{t}=v_{t}+e_{\mathcal{J}_{\alpha}}\hat{v}_{t},
\end{equation}
where $v_{t}$  is produced by a preexisting controller and $\hat{v}_{t}$ is produced by a low-dimensional local controller.
We assume that a preexisting controller denoted by
\begin{equation}\label{brin}
K:v_{t}=\mathcal{K}(y_{t}),
\end{equation}
where $\mathcal{K}$ denotes a controller dynamical map, has been implemented to stabilize the interconnected system $\Sigma$ in (\ref{sysd}), namely
\begin{equation}\label{brsta}
\psi_{t+1}=A+B\mathcal{K}(C\psi_{t})
\end{equation}
is internally stable.
Note that the broadcast controller (\ref{agcon}) without $u_{i}^{\prime}$ is a special case where $\mathcal{K}$ is the static map given as
\[
\mathcal{K}(y_{t})=\kappa{\rm diag}(a_{i})_{i\in \mathcal{G}}\mbox{\boldmath ${\mathit 1}$} \mbox{\boldmath ${\mathit 1}$}^{\sf T}y_{t}.
\]
Our objective here is to design a low-dimensional local controller that can improve the transient response of the preexisting control system composed of (\ref{sysd}), (\ref{resin}) and (\ref{brin}) using $\hat{v}_{t}$ in (\ref{resin}) as an input signal.

\subsection{Low-Dimensional Retrofit Controller Design Problem}

Consider a preexisting control system composed of $\Sigma$ in (\ref{sysd}) and $K$ in (\ref{brin}) under the composite input signal $u_{t}$ in (\ref{resin}), where we do not assume a particular system structure resulting from the power system example.
Then, let us formulate a design problem of a low-dimensional local controller that can suppress the propagation of the impact of $x_{0}$ belonging to the prespecified domain $\mathcal{X}_{\alpha}$ in (\ref{sysd}), which we call the \textit{local state deflection} at the $\alpha$th subsystem.
In this paper, we refer to such a low-dimensional local controller as a \textit{retrofit controller} associated with the local state deflection.
Without loss of generality, we assume that $x_{0}$ is contained in the unit ball, namely 
\begin{equation}\label{inicon}
x_{0}\in\bar{\mathcal{X}}_{\alpha},\quad\bar{\mathcal{X}}_{\alpha}:=\{x_{0}\in \mathcal{X}_{\alpha}:\|x_{0}\|\leq 1\}.
\end{equation}

We suppose that $\hat{v}_{t}$ in (\ref{resin}) is produced by the retrofit controller in the composite form of
\begin{equation}\label{piin}
\pi_{\alpha}:\hat{v}_{t}=\hat{\mathcal{K}}_{\alpha}\circ\hat{\mathcal{F}}_{\alpha}\left(e_{\mathcal{J}_{\alpha}}^{{\sf T}}y_{t},v_{t}\right)
\end{equation}
where $\hat{\mathcal{K}}_{\alpha}$ and $\hat{\mathcal{F}}_{\alpha}$ denote dynamical maps to be designed.
The meaning of the composite dynamical map in (\ref{piin}) is explained as follows.
First, $\hat{\mathcal{F}}_{\alpha}$ can be regarded as a dynamical compensator that performs dynamical filtration of the local output signal $e_{\mathcal{J}_{\alpha}}^{{\sf T}}y_{t}$ while measuring the input signal $v_{t}$ from the preexisting controller $K$ in (\ref{brin}).
This compensator is implemented to avoid unexpected interference between the preexisting controller $K$ and an additional controller, whose dynamical map is denoted by $\hat{\mathcal{K}}_{\alpha}$.
For the local state deflection (\ref{inicon}), the design of $\hat{\mathcal{K}}_{\alpha}$ is performed based on a low-dimensional model of (\ref{sysd}) described as
\begin{equation}\label{syslow}
\hat{\Xi}_{\alpha}:\left\{
\begin{array}{ccl}
\hat{\xi}_{t+1}&\hspace{-6pt}=&\hspace{-6pt}\hat{A}_{\alpha}\hat{\xi}_{t}+\hat{B}_{\alpha}\hat{v}_{t}\\
\hat{y}_{t}&\hspace{-6pt}=&\hspace{-6pt}\hat{C}_{\alpha}\hat{\xi}_{t},
\end{array}
\right.\quad\hat{\xi}_{0}\in\hat{\mathcal{X}}_{\alpha},
\end{equation}
where the system matrices $\hat{A}_{\alpha}$,  $\hat{B}_{\alpha}$, and $\hat{C}_{\alpha}$ as well as the initial condition domain $\hat{\mathcal{X}}_{\alpha}$ can be seen as a set of model parameters.
More specifically, $\hat{\mathcal{K}}_{\alpha}$ is designed such that the closed-loop dynamics
\begin{equation}\label{cllow}
\hat{\xi}_{t+1}=\hat{A}_{\alpha}\hat{\xi}_{t}+\hat{B}_{\alpha}\hat{\mathcal{K}}_{\alpha}(\hat{C}_{\alpha}\hat{\xi}_{t})
\end{equation}
is internally stable and the control performance criterion
\begin{equation}\label{concri}
\|\hat{\xi}_{t}\|_{l_{2}}\leq\epsilon,\quad\forall\hat{\xi}_{0}\in\hat{\mathcal{X}}_{\alpha}
\end{equation}
is satisfied with a given tolerance $\epsilon$.
We remark that one simple example of $\hat{\Xi}_{\alpha}$ in (\ref{syslow}) can be found as in (\ref{exred}), but there is a degree of freedom to find a more suitable low-dimensional model for the design of $\hat{\mathcal{K}}_{\alpha}$.
From this viewpoint, the set of $\hat{A}_{\alpha}$,  $\hat{B}_{\alpha}$,  $\hat{C}_{\alpha}$, and $\hat{\mathcal{X}}_{\alpha}$ in (\ref{syslow}) can be seen as a design parameter to construct the retrofit controller $\pi_{\alpha}$ in (\ref{piin}).

On the basis of the formulation above, we address the following retrofit controller design problem.

\begin{problem}\vspace{3pt}
Consider a preexisting control system composed of $\Sigma$ in (\ref{sysd}) and $K$ in (\ref{brin}) under the composite input signal $u_{t}$ in (\ref{resin}).
Find a retrofit controller $\pi_{\alpha}$ in (\ref{piin}) associated with the local state deflection (\ref{inicon}) such that the following specifications are satisfied:
\begin{itemize}
\item[(i)] For a low-dimensional model $\hat{\Xi}_{\alpha}$ in (\ref{syslow}), if $\hat{\mathcal{K}}_{\alpha}$ is designed such that the closed-loop dynamics (\ref{cllow}) is internally stable, then the entire closed-loop system composed of $\Sigma$,  $K$, and $\pi_{\alpha}$ is internally stable for any $K$ such that the closed-loop dynamics (\ref{brsta}) is internally stable.
\item[(ii)] If (\ref{concri}) is satisfied with a given tolerance $\epsilon$, then it follows that
\begin{equation}\label{bndJ}
\|x_{t}\|_{l_{2}}\leq\gamma_{K}\epsilon,\quad\forall x_{0}\in\bar{\mathcal{X}}_{\alpha}
\end{equation}
where $\gamma_{K}$ denotes a constant that is dependent on the design of $K$ but not dependent on the design of $\pi_{\alpha}$.
\item[(iii)] The dimensions of both $\hat{\mathcal{K}}_{\alpha}$ and $\hat{\mathcal{F}}_{\alpha}$ are less than a given tolerance number $\hat{n}$ such that $\hat{n}\leq n$.
\end{itemize}
\end{problem}\vspace{3pt}

Specification (i) is relevant to the capability that we can design the retrofit controller independently of the preexisting controller design.
More specifically, the design procedure of $\pi_{\alpha}$ does not require information on the system model of $K$ in (\ref{brsta}), while the input signal $v_{t}$ from $K$ is only used for implementation.
Specification (ii) is relevant to the transient response improvement for the local state deflection.
As designing a dynamical map $\hat{\mathcal{K}}_{\alpha}$ such that (\ref{concri}) is satisfied for a smaller tolerance $\epsilon$, we can attain transient response improvement in the sense of the upper bound (\ref{bndJ}).
Specification (iii) is relevant to reducing computational costs for the design and implementation of the retrofit controller.
Note that, even though one may be able to design a low-dimensional controller by regarding the preexisting control system as a controlled plant, the resultant low-dimensional controller does not generally satisfy Specifications (i) and (ii).
This is because the resultant low-dimensional controller should be a function of the preexisting controller, meaning that the low-dimensional controller is required to be redesigned when a preexisting controller is modified.

This retrofit control is practically reasonable in the sense that several local state deflections at respective subsystems can be handled by individual retrofit controllers, which can be predesigned independently.
In Section~\ref{secrcs}, supposing that $\mathcal{J}_{\alpha}$ is given in advance, we perform theoretical analysis to solve the retrofit controller design problem.
Then, through numerical simulation in Section~\ref{secnumex}, we investigate how the selection of  $\mathcal{J}_{\alpha}$, as well as the dimension of $\pi_{\alpha}$, affects the transient response improvement for the local state deflection.
In the numerical simulation, we show the trade-off relation between the dimension of retrofit controllers and their control performance.

\section{Retrofit Control System Design}\label{secrcs}

\subsection{Controller Design Via Hierarchical State-Space Expansion}\label{secfun}

In this subsection, we provide an overview of the control system design approach based on state-space expansion, called \textit{hierarchical state-space expansion}.
It will be found that this state-space expansion is a key to solving the retrofit controller design problem in Section~\ref{secpf}.
In the following, for simplicity of notation, we omit the subscript $\alpha$, e.g., $\pi_{\alpha}$ and $\mathcal{J}_{\alpha}$ are denoted by $\pi$ and $\mathcal{J}$, as long as there is no chance of confusion.

For the interconnected system $\Sigma$ in (\ref{sysd}) with the composite input signal $u_{t}$ in (\ref{resin}), we first show the following fact relevant to retrofit controller design.

\begin{proposition}\vspace{3pt}\label{propfun}
Let $\hat{n}$ be a natural number such that $\hat{n}\leq n$.
For a left invertible matrix $P\in \mathbb{R}^{n\times\hat{n}}$ and its left inverse $P^{\dagger}\in \mathbb{R}^{\hat{n}\times n}$, define
\begin{equation}\label{hatmat}
\hat{A}:=P^{\dagger}AP,\quad\hat{B}:=P^{\dagger}Be_{\mathcal{J}},\quad\hat{C}:=e_{\mathcal{J}}^{\sf T}CP.
\end{equation}
Consider the cascade interconnection of systems
\begin{equation}\label{syscon}
\begin{array}{ccl}
\Sigma&\hspace{-6pt}:&\hspace{-6pt}
\left\{
\begin{array}{ccl}
x_{t+1}&\hspace{-6pt}=&\hspace{-6pt}Ax_{t}+B(v_{t}+e_{\mathcal{J}}\hat{v}_{t})\\
y_{t}&\hspace{-6pt}=&\hspace{-6pt}Cx_{t}
\end{array}
\right.\vspace{3pt}\\
\hat{\Sigma}&\hspace{-6pt}:&\hspace{-6pt}
\left\{
\begin{array}{ccl}
\hat{x}_{t+1}&\hspace{-6pt}=&\hspace{-6pt}\hat{A}\hat{x}_{t}+P^{\dagger}Bv_{t}+{\mathit\Gamma}x_{t}\\
\hat{y}_{t}&\hspace{-6pt}=&\hspace{-6pt}e_{\mathcal{J}}^{\sf T}y_{t}-\hat{C}\hat{x}_{t}
\end{array}
\right.
\end{array}
\end{equation}
where  ${\mathit\Gamma}:=P^{\dagger}A-\hat{A}P^{\dagger}$.
Assume that 
\begin{equation}\label{conBC}
{\rm im}\hspace{1.5pt}Be_{\mathcal{J}}\subseteq{\rm im}\hspace{1.5pt} P,\quad
{\rm ker}\hspace{1.5pt}P^{\dagger}\subseteq{\rm ker}\hspace{1.5pt}e_{\mathcal{J}}^{\sf T}C.
\end{equation}
Then, the feedback system of $\Sigma$ and $\hat{\Sigma}$ interconnected by
\begin{equation}\label{fedcon}
K:v_{t}=\mathcal{K}(y_{t}),\quad\hat{K}:\hat{v}_{t}=\hat{\mathcal{K}}(\hat{y}_{t})
\end{equation}
is internally stable for any combination of feedback controllers $K$ and $\hat{K}$ if and only if each of the disjoint feedback systems
\begin{equation}\label{dessta}
\psi_{t+1}=A\psi_{t}+B\mathcal{K}(C\psi_{t}),\quad
\hat{\xi}_{t+1}=\hat{A}\hat{\xi}_{t}+\hat{B}\hat{\mathcal{K}}(\hat{C}\hat{\xi}_{t})
\end{equation}
is internally stable.
\end{proposition}\vspace{3pt}

\begin{IEEEproof}
The feedback system of $\Sigma$ and $\hat{\Sigma}$ in (\ref{syscon}) under the interconnection of (\ref{fedcon}) is given by
\[
\left\{
\begin{array}{ccl}
x_{t+1}&\hspace{-6pt}=&\hspace{-6pt}Ax_{t}+B\mathcal{K}(Cx_{t})+Be_{\mathcal{J}}\hat{\mathcal{K}}(e_{\mathcal{J}}^{\sf T}Cx_{t}-\hat{C}\hat{x}_{t})\vspace{0pt}\\
\hat{x}_{t+1}&\hspace{-6pt}=&\hspace{-6pt}\hat{A}\hat{x}_{t}+P^{\dagger}B\mathcal{K}(Cx_{t})+
{\mathit\Gamma}x_{t}.
\end{array}
\right.
\]
From (\ref{conBC}), it follows that
\[
P\hat{B}=Be_{\mathcal{J}},\quad\hat{C}P^{\dagger}=e_{\mathcal{J}}^{\sf T}C.
\]
Considering the coordinate transformation of 
\[
\xi_{t}=\overline{P}\ \!\overline{P}^{\dagger}x_{t}+P\hat{x}_{t},\quad\hat{\xi}_{t}=P^{\dagger}x_{t}-\hat{x}_{t},
\]
whose inverse is given by
\begin{equation}\label{defx}
x_{t}=\xi_{t}+P\hat{\xi}_{t},\quad\hat{x}_{t}=P^{\dagger}\xi_{t},
\end{equation}
where $PP^{\dagger}+\overline{P}\ \!\overline{P}^{\dagger}=I$, we have
\begin{equation}\label{clsxi}
\left\{\hspace{-1pt}
\begin{array}{ccl}
\hat{\xi}_{t+1}&\hspace{-7pt}=&\hspace{-7pt}\hat{A}\hat{\xi}_{t}+\hat{B}\hat{\mathcal{K}}(\hat{C}\hat{\xi}_{t})\\
\xi_{t+1}&\hspace{-7pt}=&\hspace{-7pt}A(\xi_{t}+P\hat{\xi}_{t})-P\hat{A}\hat{\xi}_{t}+B\mathcal{K}(C(\xi_{t}+P\hat{\xi}_{t})).
\end{array}
\right.\hspace{-1pt}
\end{equation}
This closed-loop system can be seen as the cascade system
\begin{equation}\label{hieexp}
\left\{
\begin{array}{ccl}
\hat{\xi}_{t+1}&\hspace{-6pt}=&\hspace{-6pt}\hat{A}\hat{\xi}_{t}+\hat{B}\hat{v}_{t}\\
\xi_{t+1}&\hspace{-6pt}=&\hspace{-6pt}A\xi_{t}+Bv_{t}+(AP-P\hat{A})\hat{\xi}_{t}
\end{array}
\right.
\end{equation}
with the controllers of
\begin{equation}\label{winput}
K:v_{t}=\mathcal{K}(C(\xi_{t}+P\hat{\xi}_{t})),\quad\hat{K}:\hat{v}_{t}=\hat{\mathcal{K}}(\hat{C}\hat{\xi}_{t}),
\end{equation}
which are equivalent to $K$ and $\hat{K}$ in (\ref{fedcon}).
Note that, owing to the cascade structure of (\ref{clsxi}), it is internally stable for any combination of $K$ and $\hat{K}$ if and only if both systems in (\ref{dessta}) are internally stable.
Hence, the claim is proven.
\end{IEEEproof}\vspace{3pt}

In Proposition~\ref{propfun}, the feedback controller $K$ in (\ref{fedcon}) corresponds to the preexisting  controller $K$ in (\ref{brin}) that stabilizes the interconnected system $\Sigma$ in (\ref{syscon}).
On the other hand, $\hat{K}$ in (\ref{fedcon}) can be regarded as an additional controller that stabilizes the low-dimensional model given by $\hat{A},\ \hat{B}$, and $\hat{C}$.
Note that the redundant state equation (\ref{syscon}) is equivalently transformed to the cascade state equation (\ref{hieexp}) by the coordinate transformation (\ref{defx}).
The essence of the state-space expansion is that the sum of the states $\xi_{t}$ and $\hat{\xi}_{t}$ in (\ref{hieexp}) coincides with the state $x_{t}$ in (\ref{syscon}).
From this viewpoint, the control system design for the transformed dynamics (\ref{hieexp}) makes sense also for the actual dynamics $\Sigma$ in (\ref{sysd}).
We refer to this state-space expansion of $\Sigma$, which yields the cascade system (\ref{hieexp}), as the hierarchical state-space expansion.

\begin{figure}[t]
\begin{center}
\includegraphics[width=85mm]{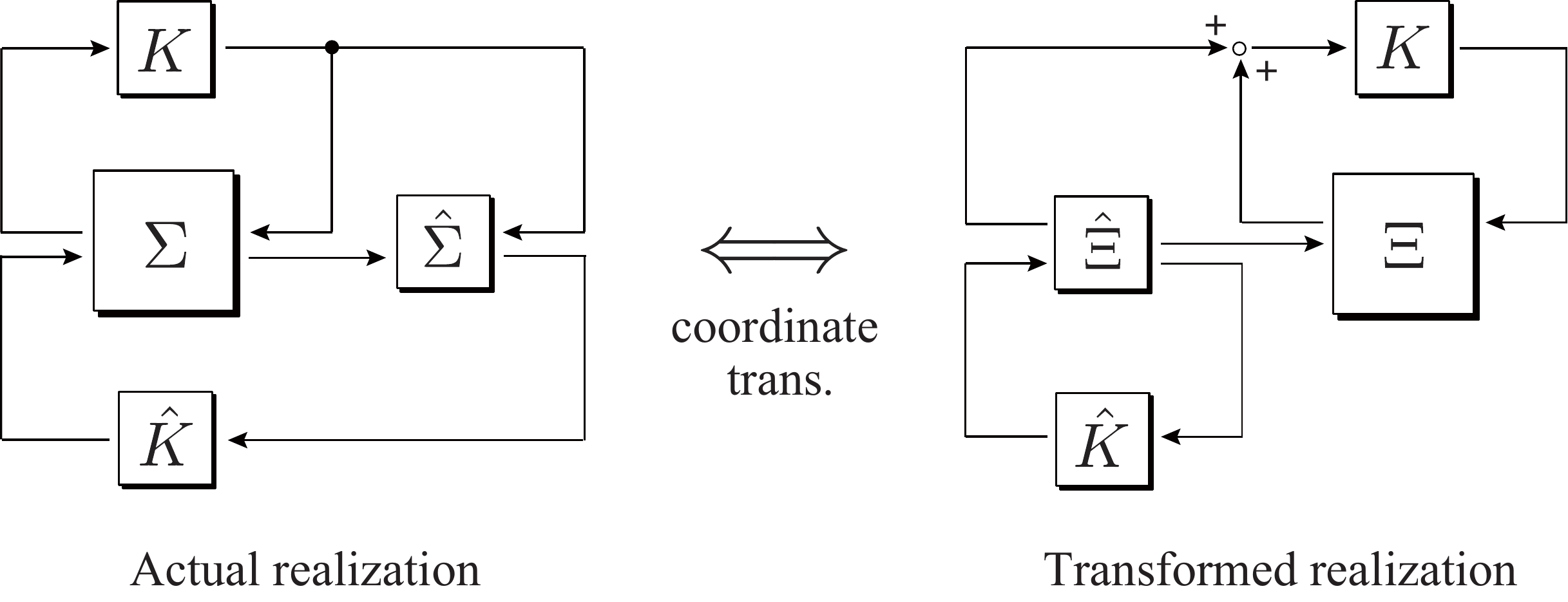}
\end{center}
\vspace{-3pt}
\caption{Signal-flow diagrams of actual and transformed dynamics.}
\label{figlembd}
\end{figure}

The relation between the actual and transformed systems with the controllers $K$ and $\hat{K}$ in (\ref{fedcon}) is depicted in Fig.~\ref{figlembd}, where the dynamics of $\hat{\xi}_{t}$ and $\xi_{t}$ in (\ref{hieexp}) are denoted by $\hat{\Xi}$ and $\Xi$, respectively.
From this figure, we see that $\hat{\Sigma}$ in (\ref{syscon}) can be regarded as a compensator that performs dynamical filtration of the output signal $y_{t}$ sent to the additional controller $\hat{K}$.
Because the feedback system on the right of Fig.~\ref{figlembd} is composed of the cascade of two feedback systems, stability analysis and control performance analysis can be performed in a systematic manner.
In this sense, the hierarchical state-space expansion has good compatibility with the retrofit controller design.

It should be noted that, for the implementation of the compensator $\hat{\Sigma}$ in (\ref{syscon}), we require an additional output signal ${\mathit\Gamma}x_{t}$ measured from the interconnected system $\Sigma$, unless $P=I$ that leads to ${\mathit\Gamma}=0$.
In general, such a particular signal is not available due to the limitations of practical sensor allocations.
To eliminate this unrealistic hypothesis, we show the following fact based on unobservable subspace matching, attained by selecting $P$ as being compatible with the output ports of $\mathcal{J}$.

\begin{proposition}\vspace{3pt}\label{propkry}
Consider the compensator $\hat{\Sigma}$ in (\ref{syscon}) and let $\hat{y}^{\prime}_{t}$ denote the output signal $\hat{y}_{t}$ when $x_{t}=0$  is imposed for all $t\geq 0$.
For a natural number $\tau$, if
\begin{equation}\label{conC}
{\rm ker}\hspace{1.5pt}P^{\dagger}\subseteq{\rm ker}
\left[\begin{array}{c}
e_{\mathcal{J}}^{{\sf T}}C\\
e_{\mathcal{J}}^{{\sf T}}CA\\
\vdots\\
e_{\mathcal{J}}^{{\sf T}}CA^{\tau-1}
\end{array}
\right],
\end{equation}
then it follows that
\begin{equation}\label{yid}
\hat{y}_{t}=\hat{y}^{\prime}_{t},\quad t=0,1,\ldots\tau-1
\end{equation}
for any sequences of $v_{t}$,  $x_{t}$, and $y_{t}$.
\end{proposition}\vspace{3pt}

\begin{IEEEproof}
From (\ref{conC}), we see that
\[
e_{\mathcal{J}}^{{\sf T}}CA^{k}PP^{\dagger}=e_{\mathcal{J}}^{{\sf T}}CA^{k},\quad k=0,1,\ldots,\tau-1.
\]
Using this relation iteratively, we have
\[
\sum_{k=1}^{t}\hat{C}\hat{A}^{k-1}{\mathit\Gamma}x_{t-k}=\sum_{k=1}^{t}(e_{\mathcal{J}}^{{\sf T}}CA^{k}-e_{\mathcal{J}}^{{\sf T}}CA^{k})x_{t-k}=0
\]
for all $t=0,1,\ldots,\tau-1$.
This implies that the input-to-output map of $\hat{\Sigma}$ from $x_{t}$ to $\hat{C}\hat{x}_{t}$ is zero within the finite-time interval.
Thus, the claim follows.
\end{IEEEproof}\vspace{3pt}

Proposition~\ref{propkry} shows that, if we select $P$ such that (\ref{conC}) holds, then the implementation of the compensator $\hat{\Sigma}$ in (\ref{syscon}) does not require the additional output signal ${\mathit\Gamma}x_{t}$ within the finite-time interval, as in (\ref{yid}).
Note that (\ref{conC}) represents the condition of unobservable subspace matching, which leads to
\[
\hat{C}\hat{x}_{t}=e_{\mathcal{J}}^{{\sf T}}Cx_{t},\quad t=0,1,\ldots,\tau-1,
\]
where each $\hat{x}_{t}$ and $x_{t}$ obeys the dynamics of
\[
\hat{x}_{t+1}=\hat{A}\hat{x}_{t}+P^{\dagger}Bv_{t},\quad x_{t+1}=Ax_{t}+Bv_{t}
\]
with $\hat{x}_{0}=P^{\dagger}x_{0}$ and a sequence of $v_{t}$.
This output matching is crucial for observer design based on the low-dimensional model.
The value of $\tau$, which is relevant to the rank of $P$, is one of design parameters in the retrofit controller design described below.

\subsection{Retrofit Controller Design with Performance Analysis}\label{sectcd}

In this subsection, on the basis of Propositions~\ref{propfun} and \ref{propkry}, we give a solution to the retrofit controller design problem in Section~\ref{secpf}.
It will be found that the retrofit controller $\pi$ in (\ref{piin}) is obtained as the composite dynamics of $\hat{\Sigma}$ and $\hat{K}$ in Proposition~\ref{propfun}.

As shown in Proposition~\ref{propkry}, if (\ref{conC}) holds, the input-to-output map from ${\mathit\Gamma}x_{t}$ to $\hat{y}_{t}$ of $\hat{\Sigma}$ in (\ref{syscon}) is zero within the finite-time interval $[0,\tau)$. 
This implies that, during the time interval, $\hat{\Sigma}$ can be implemented as
\begin{equation}\label{compss}
\hat{\Sigma}:\left\{
\begin{array}{ccl}
\hat{x}_{t+1}&\hspace{-6pt}=&\hspace{-6pt}\hat{A}\hat{x}_{t}+P^{\dagger}Bv_{t}\\
\hat{y}_{t}&\hspace{-6pt}=&\hspace{-6pt}e_{\mathcal{J}}^{\sf T}y_{t}-\hat{C}\hat{x}_{t},
\end{array}
\right.
\end{equation}
which corresponds to a state-space realization of the dynamical map $\hat{\mathcal{F}}_{\alpha}$ in (\ref{piin}).
In order to be compatible with the time interval, let us consider giving the state-space realization of $\hat{K}$ in (\ref{fedcon}) as a switching state feedback controller that is based on a finite-time output feedback observation.
This can be described as
\begin{equation}\label{dynkh}
\hat{K}:\left\{
\begin{array}{ccl}
\hat{z}_{t+1}&\hspace{-6pt}=&\hspace{-6pt}\hat{A}\hat{z}_{t}+\hat{B}\hat{v}_{t}+\sigma_{t}\hat{H}(\hat{y}_{t}-\hat{C}\hat{z}_{t})\\
\hat{v}_{t}&\hspace{-6pt}=&\hspace{-6pt}\sigma_{t}\hat{F}\hat{z}_{t}+(1-\sigma_{t})\hat{G}\hat{z}_{t}
\end{array}
\right.
\end{equation}
where $\hat{F}$,  $\hat{G}$, and $\hat{H}$ denote feedback gains designed below, and $\sigma_{t}$ denotes the switching signal given by
\begin{equation}\label{sigma}
\sigma_{t}=\left\{
\begin{array}{cl}
1,&t=0,1,\ldots,\tau-1\\
0,&{\rm otherwise}.
\end{array}
\right.
\end{equation}
The dynamics of $\hat{z}_{t}$ aims at calibrating the observation error based on the finite-time output feedback associated with $\sigma_{t}$ in (\ref{sigma}).
The switching controller $\hat{K}$ in (\ref{dynkh}) corresponds to a state-space realization of the dynamical map $\hat{\mathcal{K}}_{\alpha}$ in (\ref{piin}).
Thus, the retrofit controller $\pi$ in (\ref{piin}) is represented as
\begin{equation}\label{dypi}
\hspace{2pt}\pi:\left\{\hspace{-2pt}
\begin{array}{ccl}
\hat{x}_{t+1}&\hspace{-8pt}=&\hspace{-8pt}\hat{A}\hat{x}_{t}+P^{\dagger}Bv_{t}\\
\hat{z}_{t+1}&\hspace{-8pt}=&\hspace{-8pt}\hat{A}\hat{z}_{t}+\hat{B}\hat{w}_{t}+\sigma_{t}\hat{H}
\bigl\{e_{\mathcal{J}}^{{\sf T}}y_{t}-\hat{C}(\hat{z}_{t}+\hat{x}_{t})\bigr\}\\
\hat{v}_{t}&\hspace{-8pt}=&\hspace{-8pt}\sigma_{t}\hat{F}\hat{z}_{t}+(1-\sigma_{t})\hat{G}\hat{z}_{t}.
\end{array}
\right.\hspace{-10pt}
\end{equation}
This can be seen as the composite dynamics of $\hat{\Sigma}$ in (\ref{compss}) and $\hat{K}$ in (\ref{dynkh}), which are both $\hat{n}$-dimensional.

For selection of the image of $P$, let us consider 
\begin{equation}\label{conB}
\mathcal{X}+{\rm im}\hspace{1.5pt}[Be_{\mathcal{J}}\ ABe_{\mathcal{J}}\ \cdots\ A^{\nu-1}Be_{\mathcal{J}}]\subseteq{\rm im}\hspace{1.5pt}P,
\end{equation}
where $\mathcal{X}$ is the available domain of the local state deflection (\ref{inicon}), and the value of $\nu$ can be regarded as another design parameter.
Note that 
\[
{\rm im}\hspace{1.5pt}[Be_{\mathcal{J}}\ ABe_{\mathcal{J}}\ \cdots\ A^{\nu-1}Be_{\mathcal{J}}]\subseteq{\rm im}\hspace{1.5pt}P,
\]
is sufficient to the left condition in (\ref{conBC}), and it corresponds to the controllable subspace matching condition, which leads to
\[
 P\hat{z}_{t}=x_{t},\quad t=0,1,\ldots,\nu,
\]
where each $\hat{z}_{t}$ and $x_{t}$ obeys the dynamics of
\[
\hat{z}_{t+1}=\hat{A}\hat{z}_{t}+\hat{B}\hat{v}_{t},\quad x_{t+1}=Ax_{t}+Be_{\mathcal{I}}\hat{v}_{t}
\]
with $\hat{z}_{0}=0,\ x_{0}=0$, and a sequence of $\hat{v}_{t}$.
This controllable subspace matching is crucial for the transient response improvement based on the low-dimensional model.
On the other hand, the inclusion
\begin{equation}\label{ascalx}
\mathcal{X}\subseteq{\rm im}\hspace{1.5pt}P,
\end{equation}
implies that, for any value of $x_{0}\in \mathcal{X}$, there exists some $\hat{\xi}_{0}$ such that
\begin{equation}\label{inidec}
x_{0}=P\hat{\xi}_{0}.
\end{equation}
This implies that $\xi_{0}=0$ in (\ref{defx}), or equivalently, the initial condition of $\hat{\Sigma}$, involved in (\ref{dypi}), is to be given as $\hat{x}_{0}=0$.
Note that $\hat{\xi}_{0}$ is supposed to be unknown.
Thus, its dynamical evolution is to be estimated by the aforementioned finite-time output feedback observer.
Generalization to the case where $\mathcal{X}\not\subseteq{\rm im}$\hspace{1.5pt}$P$ will be given in Section~\ref{secgen}.

For the switching controller $\hat{K}$ in (\ref{dynkh}), the closed-loop dynamics in the right of (\ref{dessta}) is stable if $\hat{A}$ is stable and the feedback gain $\hat{G}$ is given such that $\hat{A}+\hat{B}\hat{G}$ is stable.
Under these suppositions, Proposition~\ref{propfun} shows that the feedback system composed of the interconnected system $\Sigma$ in (\ref{sysd}), the preexisting controller $K$ in (\ref{brin}), and the retrofit controller $\pi$ in (\ref{dypi}) is internally stable, i.e., Specification (i) in the retrofit controller design problem is attained.
We remark that the choices of $\hat{F}$ and $\hat{H}$ are not dependent on the closed-loop system stability because the feedback control with these gains is switched off for $ t\geq\tau$, but they are relevant to the transient response improvement relevant to Specification (ii).

For Specification (ii), let us determine the design criteria for the feedback gains $\hat{F}$,  $\hat{G}$, and $\hat{H}$ in (\ref{dypi}).
Consider the $\hat{n}$-dimensional model $\hat{\Xi}$ in (\ref{syslow}), whose system matrices are given as in (\ref{hatmat}).
The initial condition domain $\hat{\mathcal{X}}$ is given as
\begin{equation}
\hat{\mathcal{X}}=\{P^{\dagger}x_{0}\in \mathbb{R}^{\hat{n}}:x_{0}\in\bar{\mathcal{X}}\}.
\end{equation}
We analyze the $l_{2}$-norm of $\xi_{t}$ under the implementation of the switching controller $\hat{K}$ in (\ref{dynkh}).
To this end, we suppose that the feedback gains $\hat{F}$ and $\hat{H}$ are designed such that the closed-loop dynamics
\begin{equation}\label{cldyn1}
\left[
\begin{array}{cc}
\hat{\xi}_{t+1}\\
\hat{z}_{t+1}
\end{array}
\right]
=
\left[
\begin{array}{cc}
\hat{A}&\hat{B}\hat{F}\\
\hat{H}\hat{C}&\hat{A}+\hat{B}\hat{F}-\hat{H}\hat{C}
\end{array}
\right]
\left[
\begin{array}{cc}
\hat{\xi}_{t}\\
\hat{z}_{t}
\end{array}
\right]
\end{equation}
satisfies the finite-horizon criteria of
\begin{equation}\label{spec1}
\hspace{2pt}
\begin{array}{rr}
\|\hat{\xi}_{t}-\hat{z}_{t}\|_{l_{2}[\tau-1]}\leq\gamma_{1},&\quad\hspace{-6pt}
\|\hat{\xi}_{\tau}-\hat{z}_{\tau}\|\leq\delta_{1},
\\
\|\hat{z}_{t}\|_{l_{2}[\tau-1]}\leq\gamma_{2},&\quad\hspace{-6pt}\|\hat{z}_{\tau}\|\leq\delta_{2},
\end{array}
\quad\forall\hat{\xi}_{0}\in\hat{\mathcal{X}}\hspace{-12pt}
\end{equation}
with given tolerances $\gamma_{1}$ and $\gamma_{2}$, which are relevant to the finite-time $l_{2}$-norm, and $\delta_{1}$ and $\delta_{2}$, which are relevant to the terminal states at $ t=\tau$.
On the other hand, $\hat{G}$ is designed such that the closed-loop dynamics
\begin{equation}\label{cldyn2}
\hat{z}_{t+1}^{\prime}=(\hat{A}+\hat{B}\hat{G})\hat{z}_{t}^{\prime}
\end{equation}
satisfies the infinite-horizon criterion of
\begin{equation}\label{spec2}
\|\hat{z}_{t}^{\prime}\|_{l_{2}}\leq\gamma_{3},\quad\forall\hat{z}_{0}^{\prime}\in \mathcal{U}
\end{equation}
with a given tolerance $\gamma_{3}$, where $\mathcal{U}$ denotes the unit ball.
See Section~\ref{secrems} below for a way to find desirable $\hat{F}$,  $\hat{G}$, and $\hat{H}$.
In this formulation, the following performance analysis is performed.

\begin{lemma}\label{leml2bnd}\vspace{3pt}
Consider the closed-loop system composed of the $\hat{n}$-dimensional model $\hat{\Xi}$ in (\ref{syslow}) and the switching controller 
$\hat{K}$ in (\ref{dynkh}), whose system matrices are given as in (\ref{hatmat}).
If the feedback gains $\hat{F}$,  $\hat{G}$, and $\hat{H}$ in (\ref{dypi}) are designed such that (\ref{spec1}) and (\ref{spec2}) hold, then (\ref{concri}) is satisfied with 
\begin{equation}\label{epsdef}
\epsilon=\sqrt{(\gamma_{1}+\gamma_{2})^{2}+(\sqrt{q_{0}}\delta_{1}+\gamma_{3}\delta_{2})^{2}}
\end{equation}
where $q_{0}>0$ denotes the maximal eigenvalue of the positive definite matrix $Q$ such that
\begin{equation}\label{obsG}
\hat{A}^{\sf T}Q\hat{A}+I=Q.
\end{equation}
\end{lemma}\vspace{3pt}

\begin{IEEEproof}
Let $\hat{e}_{t}:=\hat{\xi}_{t}-\hat{z}_{t}$ denote the observation error.
The closed-loop system of interest is equivalent to
\begin{equation}\label{obseq1}
\left[\hspace{-2pt}
\begin{array}{cc}
\hat{e}_{t+1}\\
\hat{z}_{t+1}
\end{array}
\hspace{-2pt}
\right]
=
\left[\hspace{-2pt}
\begin{array}{cc}
\hat{A}-\hat{H}\hat{C}&\hspace{-2pt}0\\
\hat{H}\hat{C}&\hspace{-2pt}\hat{A}+\hat{B}\hat{F}
\end{array}
\hspace{-2pt}\right]
\left[\hspace{-2pt}
\begin{array}{cc}
\hat{e}_{t}\\
\hat{z}_{t}
\end{array}
\hspace{-2pt}\right]
\end{equation}
during the time interval of $t\in[0,\tau)$ and
\begin{equation}\label{obseq2}
\left[\hspace{-2pt}
\begin{array}{cc}
\hat{e}_{t+1}\\
\hat{z}_{t+1}
\end{array}
\hspace{-2pt}
\right]
=
\left[\hspace{-2pt}
\begin{array}{cc}
\hat{A}&\hspace{-2pt}0\\
0&\hspace{-2pt}\hat{A}+\hat{B}\hat{G}
\end{array}
\hspace{-2pt}\right]
\left[\hspace{-2pt}
\begin{array}{cc}
\hat{e}_{t}\\
\hat{z}_{t}
\end{array}
\hspace{-2pt}\right]
\end{equation}
for $ t\geq\tau$.
Note that the $l_{2}$-norm of $\hat{\xi}_{t}$ is decomposed as
\[
\|\hat{\xi}_{t}\|_{l_{2}}^{2}=\sum_{t=0}^{\tau-1}\|\hat{\xi}_{t}\|^{2}+\sum_{t=\tau}^{\infty}\|\hat{\xi}_{t}\|^{2}=\|\hat{\xi}_{t}\|_{l_{2}[\tau-1]}^{2}+\|\hat{\xi}_{t}^{\prime}\|_{l_{2}}^{2}
\]
where $\hat{\xi}_{t}^{\prime}:=\hat{\xi}_{t+\tau}$.
For the first term, we see from (\ref{spec1}) that
\[
\|\hat{\xi}_{t}\|_{l_{2}[\tau-1]}=\|(\hat{\xi}_{t}-\hat{z}_{t})+\hat{z}_{t}\|_{l_{2}[\tau-1]}\leq\gamma_{1}+\gamma_{2},\quad\forall\hat{\xi}_{0}\in\hat{\mathcal{X}}.
\]
In a similar manner, the second term is bounded as
\[
\|\hat{\xi}_{t}^{\prime}\|_{l_{2}}\leq\|\hat{e}_{t}^{\prime}\|_{l_{2}}+\|\hat{z}_{t}^{\prime}\|_{l_{2}}
\]
where $\hat{e}_{t}^{\prime}:=\hat{e}_{t+\tau}$ and $\hat{z}_{t}^{\prime}:=\hat{z}_{t+\tau}$.
Note that the norm of $\hat{z}_{0}^{\prime}=\hat{z}_{\tau}$ is bounded as in (\ref{spec1}).
Thus, (\ref{spec2}) implies that
\[
\|\hat{z}_{t}^{\prime}\|_{l_{2}}\leq\|\hat{z}_{\tau}\|\sup_{\hat{z}_{0}^{\prime}\in \mathcal{U}}\|\hat{z}_{t}^{\prime}\|_{l_{2}}=\gamma_{3}\delta_{2},\quad\forall\hat{\xi}_{0}\in\hat{\mathcal{X}}.
\]
On the other hand, for the observability Gramian $Q$ in (\ref{obsG}) associated with the pair $(I,\hat{A})$, we see that
\[
\sup_{\hat{e}_{0}^{\prime}\in \mathcal{U}}\|\hat{e}_{t}^{\prime}\|_{l_{2}}^{2}=
\sup_{\hat{e}_{0}^{\prime}\in \mathcal{U}}\left\{(\hat{e}_{0}^{\prime})^{{\sf T}}Q\hat{e}_{0}^{\prime}\right\}=q_{0}.
\]
Because the norm of $\hat{e}_{0}^{\prime}=\hat{\xi}_{\tau}-\hat{z}_{\tau}$ is bounded as in (\ref{spec1}), we have
\[
\|\hat{e}_{t}^{\prime}\|_{l_{2}}^{2}=\|\hat{\xi}_{\tau}-\hat{z}_{\tau}\|\sup_{\hat{e}_{0}^{\prime}\in \mathcal{U}}\|\hat{e}_{t}^{\prime}\|_{l_{2}}\leq\sqrt{q_{0}}\delta_{1},\quad\forall\hat{\xi}_{0}\in\hat{\mathcal{X}}.
\]
This proves the claim.
\end{IEEEproof}\vspace{3pt}

Lemma~\ref{leml2bnd} shows that the upper bound $\epsilon$ in (\ref{concri}) can be found as a monotone increasing function of the upper bound values $\gamma_{1}$,  $\gamma_{2}$,  $\gamma_{3}$,  $\delta_{1}$, and $\delta_{2}$ in (\ref{spec1}) and (\ref{spec2}).
This implies that, if we design the feedback gains $\hat{F}$,  $\hat{G}$, and $\hat{H}$ as decreasing these upper bound values, then we can decrease the resultant value of $\epsilon$ in the sense of (\ref{epsdef}).
Note that the observer initial condition $\hat{z}_{0}$ can be fixed as an arbitrary value, which corresponds to an initial guess of $\hat{\xi}_{0}$.
The dynamics of $\hat{z}_{t}$ evolves as decreasing the observation error $\hat{e}_{t}$ in (\ref{obseq1}) and (\ref{obseq2}) based on the finite-time output feedback of $e_{\mathcal{J}}^{{\sf T}}y_{t}$.
In particular, provided that $\hat{z}_{0}=\hat{\xi}_{0}$, (\ref{obseq1}) and (\ref{obseq2}) are reduced to the state-feedback system
\begin{equation}\label{stfbc}
\hat{z}_{t+1}=(\hat{A}+\hat{B}\hat{F})\hat{z}_{t},\quad\hat{e}_{t}=0
\end{equation}
where $\hat{F}=\hat{G}$ is supposed.
Thus, the value $\epsilon$ in (\ref{epsdef}) can be replaced with the upper bound value of
\[
\|\hat{z}_{t}^{\prime}\|_{l_{2}}\leq\epsilon,\quad\forall\hat{z}_{0}\in\hat{\mathcal{X}}
\]
for the state-feedback system (\ref{stfbc}).

For simplicity, let us assume that the preexisting controller $K$ in (\ref{brin}) is given as a static controller denoted by
\begin{equation}\label{staasmp}
K:v_{t}=Fy_{t},
\end{equation}
which guarantees the stability of
\begin{equation}\label{defAk}
A_{K}:=A+BFC.
\end{equation}
Generalization to the case of dynamical preexisting controllers will be discussed in Section~\ref{secgen}.
Then, we state the following theorem  relevant to Specification (ii).

\begin{theorem}\vspace{3pt}\label{thmprf}
Under the composite input signal $u_{t}$ in (\ref{resin}), consider the entire closed-loop system composed of the interconnected system $\Sigma$ in (\ref{sysd}), the preexisting static controller $K$ in (\ref{staasmp}) and the retrofit controller $\pi$ in (\ref{dypi}).
If the feedback gains $\hat{F}$,  $\hat{G}$, and $\hat{H}$ in (\ref{dypi}) are designed such that (\ref{spec1}) and (\ref{spec2}) hold, then (\ref{bndJ}) holds with $\epsilon$ in (\ref{epsdef}) and 
\begin{equation}\label{constant}
\gamma_{K}:=\bigl\|
(zI-A_{K})^{-1}(A_{K}-PP^{\dagger}A)P+P
\bigr\|_{h_{\infty}}
\end{equation}
where $A_{K}$ is defined as in (\ref{defAk}).
\end{theorem}\vspace{3pt}

\begin{IEEEproof}
From (\ref{defx}), we see that
\begin{equation}\label{xl2}
\|x_{t}\|_{l_{2}}=\|\xi_{t}+P\hat{\xi}_{t}\|_{l_{2}}=\|\mathcal{H}(\hat{\xi}_{t})+P\hat{\xi}_{t}\|_{l_{2}}
\end{equation}
where the dynamical map $\mathcal{H}$ corresponds to the dynamics of $\xi_{t}$ in (\ref{hieexp}) with the control input $v_{t}$ in (\ref{winput}), given by
\[
\xi_{t+1}=A_{K}\xi_{t}+(A_{K}-PP^{\dagger}A)P\hat{\xi}_{t},\quad\xi_{0}=0.
\]
Considering the $z$-transformation of $x_{t}$, we have
\begin{equation}\label{disdes}
\|x_{t}\|_{l_{2}}=\|W_{K}(z)\hat{\xi}(z)\|_{h_{2}}\leq\|W_{K}(z)\|_{h_{\infty}}\|\hat{\xi}(z)\|_{h_{2}},
\end{equation}
where $\hat{\xi}(z)$ denotes the $z$-transform of $\hat{\xi}_{t}$ and
\begin{equation}\label{wsysz}
W_{K}(z):=(zI-A_{K})^{-1}(A_{K}-PP^{\dagger}A)P+P.
\end{equation}
From Lemma~\ref{leml2bnd}, we see that
\[
\|\hat{\xi}(z)\|_{h_{2}}=\|\hat{\xi}_{t}\|_{l_{2}}\leq\epsilon,\quad\forall\hat{\xi}_{0}\in\hat{\mathcal{X}}.
\]
Thus, (\ref{xl2}) is bounded as in (\ref{bndJ}).
\end{IEEEproof}\vspace{3pt}

In Theorem~\ref{thmprf}, we see that $\gamma_{K}$ in (\ref{constant}) is independent of the feedback gains $\hat{F}$,  $\hat{G}$, and $\hat{H}$.
Thus, as long as we appropriately tune the feedback gains such that $\epsilon$ in (\ref{epsdef}) is decreased, the transient response improvement for the local state deflection is achieved in the sense of the bound (\ref{bndJ}).
Our systematic performance analysis is fully reliant on the hierarchical state-space expansion in Section~\ref{secfun}.

\begin{figure}[t]
\begin{center}
\includegraphics[width=85mm]{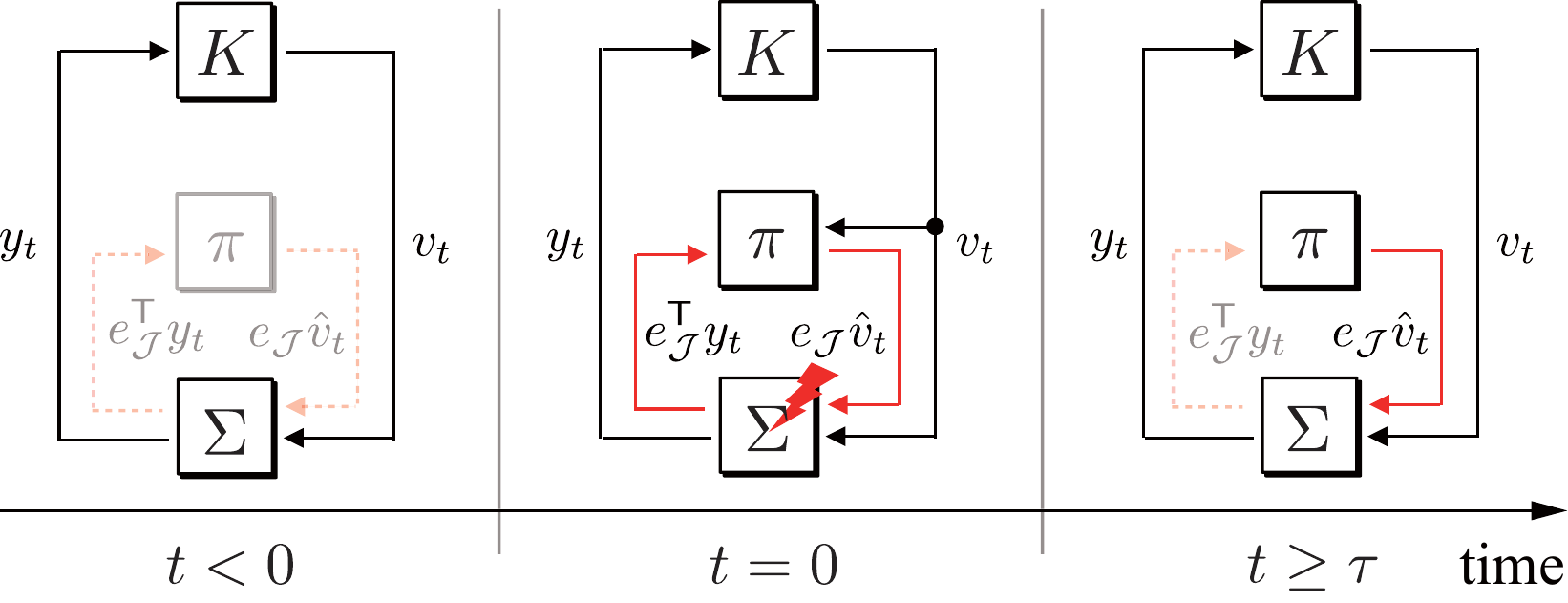}
\end{center}
\vspace{-3pt}
\caption{Sequential implementation procedure of retrofit control.}
\label{figcontrolflow}
\end{figure}

Notice that the compensation signal $\hat{C}\hat{x}_{t}$ in the dynamics of $\hat{z}_{t}$ in (\ref{dypi}) is used only within the finite-time interval associated with $\sigma_{t}$.
Thus, the computation of the dynamical evolution of $\hat{x}_{t}$, i.e., the implementation of $\hat{\Sigma}$ in (\ref{compss}), can be removed after $ t=\tau$.
In this sense, the compensator $\hat{\Sigma}$ in (\ref{compss}), involved in the retrofit controller $\pi$ in (\ref{dypi}), can be regarded as a temporal memory for dynamical filtration of $e_{\mathcal{J}}^{{\sf T}}y_{t}$.
Fig.~\ref{figcontrolflow} depicts the sequential implementation procedure of $\pi$.

Let us remark on transient response improvement in comparison with the preexisting control system before the retrofit control.
For simplicity, let us consider the case of $\hat{z}_{0}=\hat{\xi}_{0}$, i.e., the state-feedback case of (\ref{stfbc}), which leads to
\begin{equation}\label{l2rep}
\|x_{t}\|_{l_{2}}=\bigl\|W_{K}(z)\bigl(zI-(\hat{A}+\hat{B}\hat{F})\bigr)^{-1}\hat{\xi}_{0}\bigr\|_{h_{2}},
\end{equation}
where $W_{K}$ is defined as in (\ref{wsysz}).
In this representation, it turns out that the $l_{2}$-norm of $x_{t}$ without the retrofit control corresponds to the case of $\hat{F}=0$.
Because the minimization of (\ref{l2rep}) is a standard LQR design problem, we see that there exists a feedback gain $\hat{F}$ such that the $l_{2}$-norm performance specification is improved (or at least a feedback gain does not make it worse).
Even though, in principle, we can always find a minimizer $\hat{F}$ as long as we know  $W_{K}$, the minimizer depends on the parameters of the preexisting control system.
To avoid redesigning $\pi$ according to the modification of $K$, it would be reasonable to suppress the upper bound as in (\ref{disdes}).
The design scheme of $\pi$ corresponds to an extreme case where $W_{K}$ is supposed to be the all-pass system $\gamma_{K}I$.

\subsection{Remarks on Implementation}\label{secrems}

We provide several remarks on the implementation of retrofit controllers.
The first is relevant to the trade-off relation between the dimension of retrofit controllers and their control performance.
The rank of $P$, which is identical to the dimensions of $\hat{\Sigma}$ in (\ref{compss}) and $\hat{K}$ in (\ref{dynkh}), can be regarded as a design criterion to regulate the degree of transient response improvement.
Indeed, as seen from (\ref{conC}), the value of $\tau$ is necessarily less than the rank of $P$.
Recall that $\tau$ corresponds to the width of the time interval, within which the specified output signal $e_{\mathcal{J}}^{{\sf T}}y_{t}$ is fedback to the retrofit controller in order to decrease the observation error $\hat{\xi}_{t}-\hat{z}_{t}$.
As a dual argument, the rank of $P$ is also relevant to the value of $\nu$ in (\ref{conB}), which  corresponds to the dimension of the subspace that is controllable by the retrofit controller.
Thus, there is a trade-off relation between the dimension and the control  performance, which will be shown by numerical simulation in Section~\ref{secnumex}.

Next, we give a remark on finding feedback gains $\hat{F}$,  $\hat{G}$, and $\hat{H}$ such that the design criteria (\ref{spec1}) and (\ref{spec2}) are satisfied.
The problem of finding $\hat{G}$ corresponds to a standard LQR design problem, whose optimal solution can be found via a convex program, e.g., solving a system of linear matrix inequalities \cite{boyd1987linear}.
On the other hand, because the problem of finding $\hat{F}$ and $\hat{H}$ corresponds to a finite-horizon control problem, to find an optimal solution is not very simple, but a sufficient solution can be found by applying an existing method such as in \cite{ferrante2013generalised,mayne2006robust,bemporad2000output}.
Note that, even though these existing methods may produce time-variant (or state-dependent) feedback gains, generalization of $\hat{K}$ in (\ref{dynkh}) to such a time-variant feedback controller is straightforward because Proposition~\ref{propfun} is valid for any dynamical controller $\hat{K}$ stabilizing the closed-loop system in the right of (\ref{dessta}).
More generally, the dynamics of $\hat{K}$ in (\ref{dynkh}) during the time interval $[0,\tau)$ can be replaced with any observer-based feedback controller 
\[
\left\{
\begin{array}{ccl}
\hat{z}_{t+1}&\hspace{-6pt}=&\hspace{-6pt}\hat{f}_{t}(\hat{z}_{t},\hat{y}_{t})\\
\hat{v}_{t}&\hspace{-6pt}=&\hspace{-6pt}\hat{g}_{t}(\hat{z}_{t})
\end{array}
\right.
\]
such that the criteria of (\ref{spec1}) are satisfied for the $\hat{n}$-dimensional model $\hat{\Xi}$ in (\ref{syslow}).

For an algorithm to find $P$ such that (\ref{conC}) and (\ref{conB}) hold, the following biconjugation process can be used.
Denote some desirable coordinates by the sets of vectors $u_{i}$ and $v_{i}$, which are given in advance.
Our objective here is to find $P$ such that
\[
{\rm im}\hspace{1.5pt} P={\rm span}\{u_{1},\ldots,u_{k}\},\quad
{\rm ker}\hspace{1.5pt}P^{\dagger}={\rm span}\{v_{1},\ldots,v_{k}\}^{\perp},
\]
where $\perp$ indicates the orthogonal complement.
To this end, we consider the biconjugation process given by
\begin{equation}\label{wedpro}
 p_{i}:=u_{i}-\displaystyle \sum_{j=1}^{i-1}\frac{u_{i}^{\sf T}q_{j}}{p_{j}^{\sf T}q_{j}}p_{j},\quad
q_{i}:=v_{i}-\displaystyle \sum_{j=1}^{i-1}\frac{p_{j}^{\sf T}v_{i}}{p_{j}^{\sf T}q_{j}}q_{j},
\end{equation}
for which we give $p_{1}:=u_{1}$ and $q_{1}:=v_{1}$.
In \cite{chu1995rank}, it is shown that $p_{i}^{\sf T}q_{j}=0$ holds for all $i\neq j$, or equivalently
\[
Q^{{\sf T}}P=D,\quad D:={\rm diag}(p_{1}^{\sf T}q_{1},\ldots,p_{k}^{\sf T}q_{k})
\]
where $P:=[p_{1}$\ $\cdots$\ $p_{k}]$ and $Q:=[q_{1}$\ $\cdots\ q_{k}]$.
Furthermore, it follows that
\[
{\rm im}\hspace{1.5pt} P={\rm span}\{u_{1},\ldots,u_{k}\},\quad
{\rm im}\hspace{1.5pt}Q={\rm span}\{v_{1},\ldots,v_{k}\}.
\]
Thus, this leads to $P$ of interest, whose left inverse is given as $P^{\dagger}=D^{-1}Q^{{\sf T}}$.

This biconjugation process is closely related to the two-sided Lanczos procedure in the Krylov projection \cite{antoulas2005approximation}, which is used to compute the eigenvalues of large matrices as well as for model reduction.
Unfortunately, due to the fact that the eigenvalues of $\hat{A}$ in (\ref{hatmat}) are uniquely determined by the selection of the image of $P$ and the kernel of $P^{\dagger}$, the process in (\ref{wedpro}) does not necessarily produce a stable matrix $\hat{A}$, whose stability has been assumed in the retrofit controller design.
To resolve this difficulty, as long as $\Sigma$ in (\ref{sysd}) is originally stable, it can be expected that a stable approximant would be obtained when we increase the dimension of $\hat{A}$, because the eigenvalue distribution of $\hat{A}$ tends to approximate that of $A$.
The validity of this expectation will be demonstrated in Section~\ref{secnumex} numerically.
Devising a systematic way to find a stable minimal approximant is currently under investigation.

\subsection{Generalization}\label{secgen}

We provide several guidelines for generalizing our retrofit controller design method.
In Section~\ref{sectcd}, we have assumed that the domain of the local state deflection, denoted by $\mathcal{X}$, satisfies the inclusive relation in (\ref{ascalx}), which allows the  decomposition in (\ref{inidec}).
To relax this assumption, let us consider the case of $\mathcal{X}\not\subseteq{\rm im}$\hspace{1.5pt}$P$.
Note that, even in this case, the stability of the entire feedback system is still guaranteed because the closed-loop system is made internally stable.
In addition, control performance analysis alternative to Theorem~\ref{thmprf} can be carried out in a similar manner.
This is explained as follows.
Because the sum of the images of $P$ and $\overline{P}$ covers the whole space, there exist some $\hat{\xi}_{0}$ and $\hat{\xi}_{0}^{\prime}$ such that
\[
x_{0}=P\hat{\xi}_{0}+\overline{P}\hat{\xi}_{0}^{\prime}
\]
for any value of $x_{0}$.
This corresponds to a generalized version of (\ref{inidec}).
Then, we can obtain a bound alternative to (\ref{bndJ}) as 
\[
\|x_{t}\|_{l_{2}}\leq\|A_{K}^{t}\overline{P}\hat{\xi}_{0}^{\prime}\|_{l_{2}}+\gamma_{K}\epsilon,\quad\forall x_{0}\in\bar{\mathcal{X}}.
\]
This implies that the state of $\overline{P}\hat{\xi}_{0}^{\prime}$, i.e., the component of $x_{0}$ lying in the image of $\overline{P}$, can never be controlled by the retrofit controller $\pi$ in (\ref{dypi}), whose controllable subspace is determined by the image of $P$.
Therefore, to improve the transient response for the local state deflection, it is desirable that the norm of $\overline{P}\hat{\xi}_{0}^{\prime}$ be as small as possible, or equivalently, that $\mathcal{X}$ is covered by the image of $P$ as much as possible.
Note that the image of $P$ is a function of the selection of $\mathcal{J}$ as shown in (\ref{conB}).
Similar to this, the selection of $\mathcal{J}$ is also relevant to decreasing the observation error $\hat{\xi}_{t}-\hat{z}_{t}$.
From this viewpoint, we can see that the explicit association of the domain $\mathcal{X}$ with the selection of $\mathcal{J}$ is essential to improving the control performance.
This aspect will also be demonstrated in Section~\ref{secnumex}.
We remark that the theoretical analysis above can be simply generalized to the case where the input ports and the output ports are not identical.

Furthermore, in Theorem~\ref{thmprf}, we have assumed that the preexisting controller is static and its sampling and holding times are equal to the sampling time of $\Sigma$, denoted by $\Delta t$.
To relax these assumptions, notice that Proposition~\ref{propfun} is valid for any dynamical controller $K$, regardless of its dimension, its sampling and holding times, and other details.
Thus, we can straightforwardly generalize the arguments in Theorem~\ref{thmprf} to the case where a preexisting controller is dynamical and its sampling and holding times are larger than $\Delta t$.
More specifically, by denoting the sampling time by $m\Delta t$, a dynamical preexisting controller can be described as
\begin{equation}\label{dynK}
K:\left\{
\begin{array}{ccl}
\eta_{m(t+1)}&\hspace{-6pt}=&\hspace{-6pt}G\eta_{mt}+H y_{mt}\\
w_{mt}&\hspace{-6pt}=&\hspace{-6pt}F\eta_{mt}
\end{array}
\right.
\end{equation}
whose input holding is represented as
\[
v_{mt}=v_{mt+1}=\cdots=v_{m(t+1)-1}=w_{mt}.
\]
The controller parameters are designed such that
\[
\left[
\begin{array}{cc}
A^{m}&\left[B\ AB\ \cdots\ A^{m-1}B\right]F\\
HC&G
\end{array}
\right]
\]
is stable.
In fact, replacing the value of $\gamma_{K}$ in (\ref{constant}) with that corresponding to the dynamical version of $K$ in (\ref{dynK}), we can perform control performance analysis similar to that in Theorem~\ref{thmprf}.

Finally, let us consider the case where two different retrofit controllers, denoted as $\pi_{\alpha}$ and $\pi_{\beta}$, are simultaneously implemented to the preexisting control system.
The resultant composite input signal, corresponding to a generalized version of (\ref{resin}), can be represented as
\[
u_{t}=v_{t}+e_{\mathcal{J}_{\alpha}}\hat{v}_{t}+e_{\mathcal{J}_{\beta}}\hat{v}_{t}^{\prime},
\]
where $\hat{v}_{t}$ and $\hat{v}_{t}^{\prime}$ denote input signals from $\pi_{\alpha}$ and $\pi_{\beta}$ to the input ports associated with $\mathcal{J}_{\alpha}$ and $\mathcal{J}_{\beta}$, respectively.
Let us regard $\pi$ in (\ref{dypi}) as $\pi_{\alpha}$.
In this formulation, what we have to modify for $\pi_{\alpha}$ is to replace the input signal $v_{t}$ with $v_{t}+e_{\mathcal{J}_{\beta}}\hat{v}_{t}^{\prime}$ in the dynamics of $\hat{x}_{t}$.
Clearly, $\pi_{\beta}$ can be designed in a manner similar to that of $\pi_{\alpha}$.
Generalization to the case of more than two retrofit controllers can be done in the same way.

\section{Numerical Experiments}\label{secnumex}

For the power system (\ref{eprsys}) with the broadcast controller (\ref{agcon}) in Section~\ref{secmot}, we consider designing the proposed retrofit controller to improve the control performance for a local fault at the generator on Bus 107.
In the following discussion, we compare the response of the retrofit control system by varying the input and output ports, i.e., $\mathcal{J}_{\alpha}$, as well as the dimension of the retrofit controller, i.e., the rank of $P$.
In particular, for comparison with regard to the allocation of input and output ports, we consider three cases: (i) one generator on Bus 107 is used, (ii) two generators on Busses 107 and 110 are used, and (iii) three generators on Busses 104, 107, and 111 are used; see Fig.~\ref{figieee118} for the locations of the specified generators.
The retrofit controller provides individual input signals to the specified generators while measuring their average frequency as an output signal.
The feedback gains $\hat{F}=\hat{G}$ and $\hat{H}$ in (\ref{dynkh}) are designed by the LQR design technique, in which we consider minimizing a quadratic cost with respect to $P\hat{\xi}_{t}$, which corresponds to the state variable associated with the local state deflection.

\begin{figure}[t]
\begin{center}
\includegraphics[width=70mm]{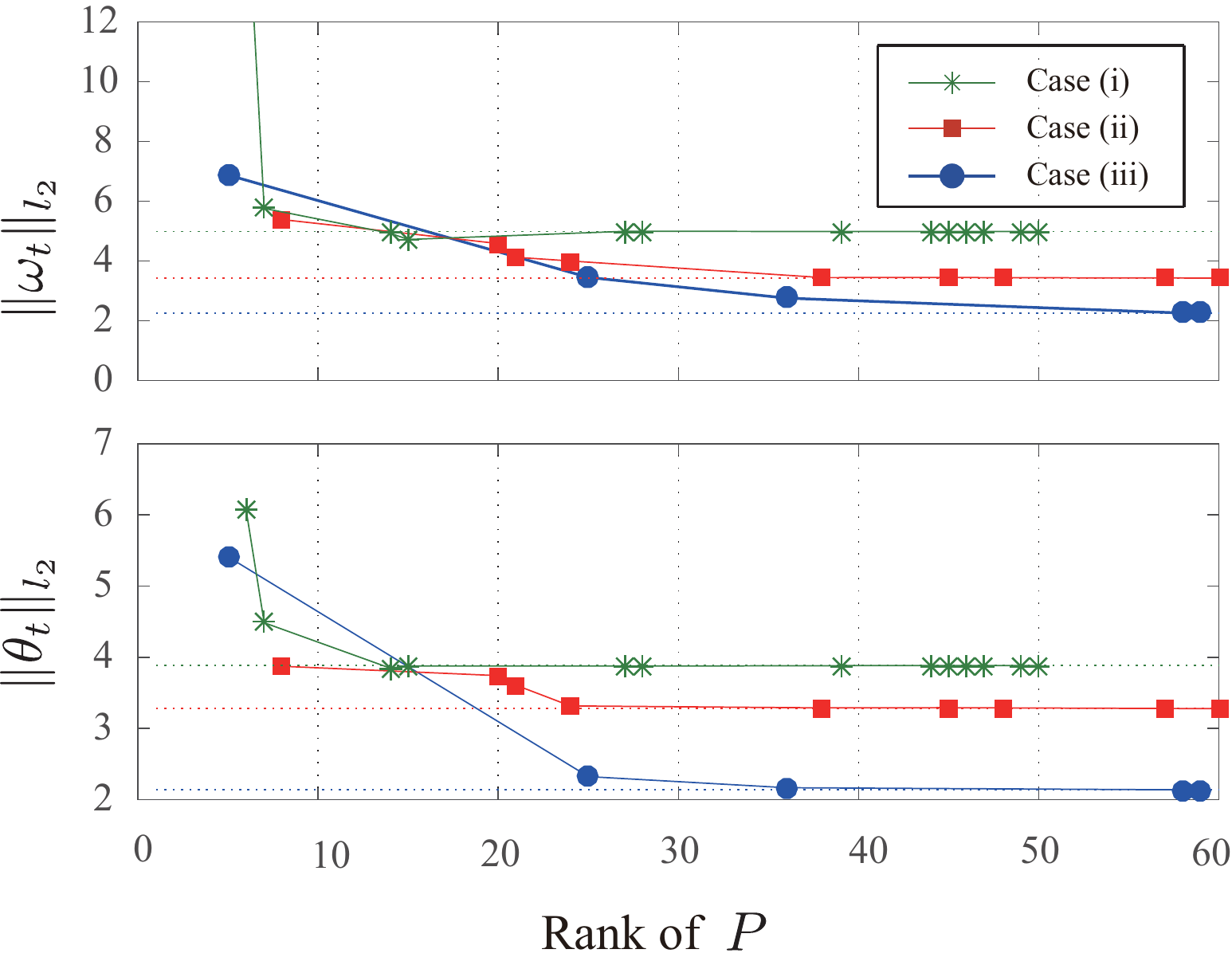}
\end{center}
\vspace{-6pt}
\caption{
The values of $\|\omega_{t}\|_{l_{2}}$ and $\|\theta_{t}\|_{l_{2}}$ versus the rank of $P$.
}
\label{figpninputp}
\end{figure}

\begin{figure*}[t]
\begin{center}
\includegraphics[width=120mm]{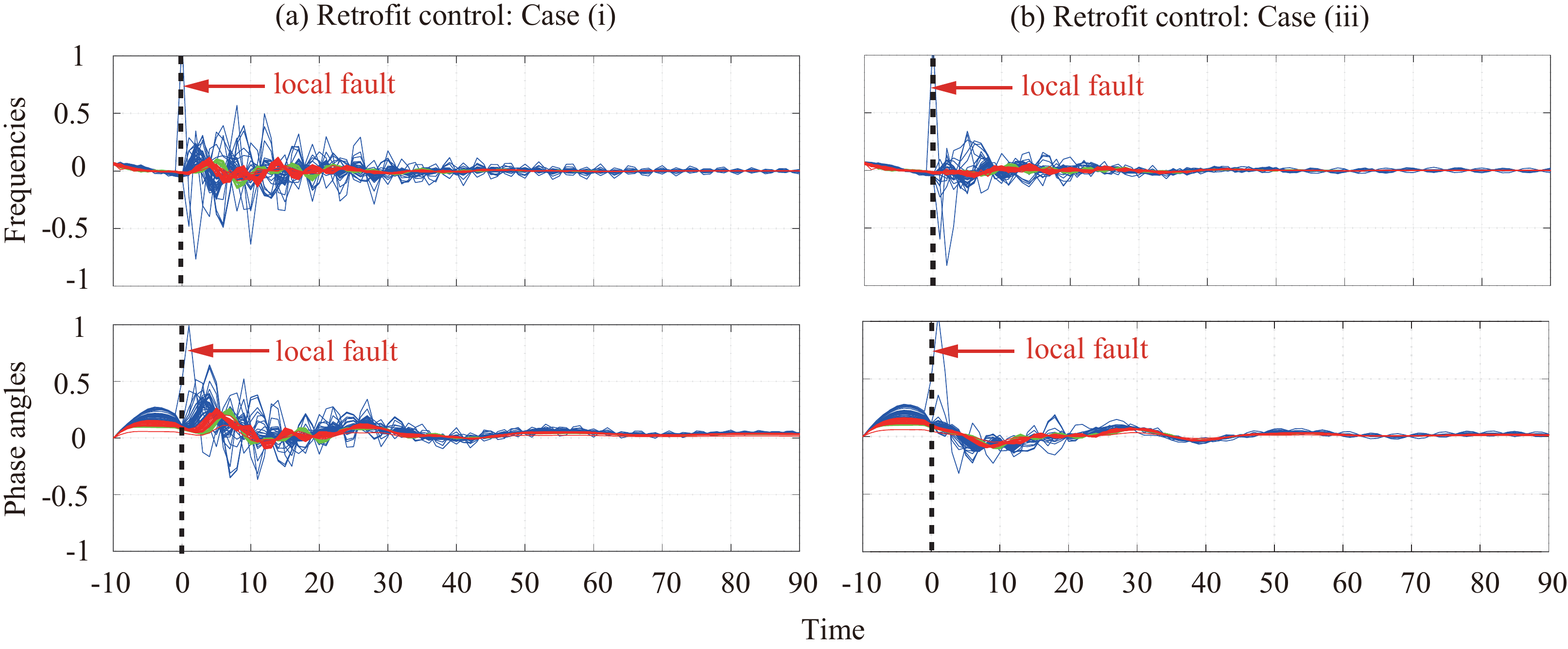}
\end{center}
\vspace{-6pt}
\caption{
System responses for local fault at generator on Bus 107.
The subfigures plot the frequencies and phase angles of all appliances whose line colors (blue, green, and red) are compatible with those of the appliance groups in Fig.~\ref{figieee118}.
}
\label{figretro}
\end{figure*}

We consider giving the initial value $\hat{z}_{0}$ in (\ref{dypi}) as
\[
\hat{z}_{0}=P^{\dagger}\tilde{x}_{0}
\]
where $\tilde{x}_{0}$ denotes a guess of the local state deflection $x_{0}$.
Note that if it is possible to give the guess as $\tilde{x}_{0}=x_{0}$, then we have $\hat{z}_{0}=\hat{\xi}_{0}$, which corresponds to the ideal situation where complete information of the state deflection is available.
Complying with the supposition that only the frequency of the generators is measurable, we associate the elements of $\tilde{x}_{0}$ only with the generator frequency identical to those of $x_{0}$.
On the other hand, because the elements of $\tilde{x}_{0}$ associated with the phase angles are not measurable, they are supposed to be zero.
Based on this, we simulate a situation of $\hat{z}_{0}\neq\hat{\xi}_{0}$.
The observation error $\hat{\xi}_{t}-\hat{z}_{t}$ is to be dynamically decreased by the finite-time output feedback of $e_{\mathcal{J}}^{{\sf T}}y_{t}$ in (\ref{dypi}).

In Fig.~\ref{figpninputp}, we plot the values of $\|\omega_{t}\|_{l_{2}}$ and $\|\theta_{t}\|_{l_{2}}$ versus the rank of $P$, which determines the dimension of the retrofit controller.
The lines with asterisks, squares, and circles correspond to the cases of (i), (ii), and (iii), respectively.
Because $\hat{A}$ in (\ref{hatmat}) is not necessarily stable for all $P$, we plot the values only when $\hat{A}$ is stable.
Furthermore, to make the comparison fair, we adjust the scales of weighting matrices in the LQR design technique such that the norms of resultant input signals are comparable.
From the figure, we find that the values of $\|\omega_{t}\|_{l_{2}}$ and $\|\theta_{t}\|_{l_{2}}$ tend to decrease, i.e., the control performance improves, as the rank of $P$ increases.
Furthermore, we see that the indices of control performance reach some particular limits, denoted by the dashed lines, which are obtained when $P$ is of the maximal rank, i.e., $P=I$.
This result suggests that we should  determine the dimension, as well as the allocation of input and output ports of retrofit controllers, while also considering the trade-off relation with respect to control performance.

For the cases of (i) and (iii), we plot the resultant system responses with the retrofit control in Figs.~\ref{figretro}(a) and (b), where we use $P$ being of rank 39 and 36 for (i) and (iii), respectively.
From these figures, we see that the transient responses improve as we increase the number of input and output ports for retrofit control.
Furthermore, the propagation of the local fault to other appliance groups are well suppressed in comparison to Fig.~\ref{figbroad}(a), where we use only the preexisting broadcast controller.
In fact, the resultant values of $\|\omega_{t}\|_{l_{2}}$ and $\|\theta_{t}\|_{l_{2}}$ are 3.88 and 4.99 in (i), and 2.19 and 2.76 in (iii), which are less than 8.67 and 15.96 found for the preexisting broadcast control.
These results demonstrate the effectiveness of our retrofit control.

\section{Concluding Remarks}

In this paper, on the basis of the hierarchical state-space expansion, we have proposed a low-dimensional retrofit controller design method for interconnected linear systems.
The proposed method is practically reasonable in the sense that a set of local faults can be handled by a set of particular retrofit controllers, which can be predesigned individually.
Furthermore, we do not need to redesign a preexisting controller that focuses on accomplishing an objective from a global viewpoint.
The efficiency of the proposed method has been shown through an example of power systems control.

In the controller design based on the hierarchical state-space expansion, a projection-based model reduction method is utilized to extract a low-dimensional model that is controlled by the retrofit controller.
Because the sizes of controllable and unobservable subspaces of the low-dimensional model increase as the model dimension increases, a trade-off relation is found between the dimension and the control performance for local state deflections.
Furthermore, the dimension and control performance are both relevant to the allocation of input and output ports for retrofit controllers.
A theoretical analysis to appropriately determine input and output port allocation, associated with individual state deflection scenarios, as well as the dimension of the retrofit controller is a future work to pursue.




\ifCLASSOPTIONcaptionsoff
  \newpage
\fi




\bibliographystyle{IEEEtran}     
\bibliography{IEEEabrv,reference,reference_CREST}

%

%





\end{document}